\documentclass{article}

\usepackage{arxiv}

\usepackage[utf8]{inputenc}
\usepackage[T1]{fontenc}

\usepackage{amsthm}  
\usepackage{newtxtext}
\usepackage{newtxmath}
\usepackage{bm}

\usepackage{algorithm}
\usepackage{algpseudocode}
\usepackage{tikz}

\usepackage{graphicx}
\usepackage{booktabs}
\usepackage{longtable}
\usepackage{mathtools}
\usepackage{hyperref}
\usepackage{multirow}

\usepackage{microtype}
\usepackage{cleveref}

\usepackage{natbib}

\newcommand{\E}{\mathbb{E}}

\newtheorem{theorem}{Theorem}[section]
\newtheorem{lemma}[theorem]{Lemma}

\newtheorem{corollary}[theorem]{Corollary}

\newtheorem{proposition}[theorem]{Proposition}

\usepackage{graphicx}
\graphicspath{{images/}} 

\allowdisplaybreaks

\title{DeePM: Regime-Robust Deep Learning for Systematic Macro Portfolio Management}
\renewcommand{\shorttitle}{DeePM: Regime-Robust Deep Learning for Systematic Macro Portfolio Management}

\author{%
  Kieran Wood\\
  Oxford-Man Institute \& \\ Machine Learning Research Group\\
  University of Oxford\\
  \texttt{kieran.wood@eng.ox.ac.uk} \\
  \AND
  Stephen J. Roberts \\
   Machine Learning Research Group\\
  University of Oxford\\
  \texttt{stephen.roberts@eng.ox.ac.uk} \\
  \And
  Stefan Zohren \\
   Machine Learning Research Group\\
  University of Oxford\\
  \texttt{stefan.zohren@eng.ox.ac.uk} \\
}

\hypersetup{
pdftitle={DeePM: Regime-Robust Deep Learning for Systematic Macro Portfolio Management},
pdfsubject={q-fin.TR, cs.LG, stat.ML},
pdfauthor={Kieran~Wood, Stephen J.~Roberts, Stefan~Zohren},
pdfkeywords={Systematic Macro, Portfolio Management, Deep Learning, Attention, Graph Neural Networks, Transaction Costs, Robust Optimization, Risk Measures},
}

\begin{document}
\maketitle
\begin{abstract}
We propose \textbf{DeePM} (\textbf{Dee}p \textbf{P}ortfolio \textbf{M}anager), a structured deep-learning macro portfolio manager trained end-to-end to maximize a robust, risk-adjusted utility. DeePM addresses three fundamental challenges in financial learning: (1) it resolves the asynchronous ``ragged filtration'' problem via a \emph{Directed Delay} (\emph{Causal Sieve}) mechanism that prioritizes causal impulse-response learning over information freshness; (2) it combats low signal-to-noise ratios via a \emph{Macroeconomic Graph Prior}, regularizing cross-asset dependence according to economic first principles; and (3) it optimizes a \emph{distributionally robust objective} where a smooth worst-window penalty serves as a differentiable proxy for Entropic Value-at-Risk (EVaR) -- a \emph{window-robust} utility encouraging strong performance in the most adverse historical subperiods.
In large-scale backtests from 2010--2025 on 50 diversified futures with highly realistic transaction costs, DeePM attains net risk-adjusted returns that are roughly twice those of classical trend-following strategies and passive benchmarks, solely using daily closing prices. Furthermore, DeePM improves upon the state-of-the-art Momentum Transformer architecture by roughly fifty percent. The model demonstrates structural resilience across the 2010s ``CTA (Commodity Trading Advisor) Winter'' and the post-2020 volatility regime shift, maintaining consistent performance through the pandemic, inflation shocks, and the subsequent higher-for-longer environment. Ablation studies confirm that strictly lagged cross-sectional attention, graph prior, principled treatment of transaction costs, and robust minimax optimization are the primary drivers of this generalization capability. 
\end{abstract}



\section{Introduction}\label{sec:intro}
The central goal of systematic portfolio management is to construct asset allocations that generalize out-of-sample under heavy-tailed returns, regime shifts, and significant trading frictions. While classical mean--variance optimization \citep{markowitz1952portfolio} provides a foundational framework, its practical deployment is plagued by ``error maximization'' \citep{michaud1989}, where small estimation errors in covariance matrices lead to unstable, turnover-intensive portfolios. Consequently, modern approaches have increasingly pivoted toward machine learning pipelines. However, most existing methods adopt a disjoint two-stage approach -- forecasting returns first, then performing a portfolio construction step -- which misaligns the training loss (Mean Squared Error) with the investor's ultimate utility (Net Risk-Adjusted Return) \citep{gu2020empirical, elmachtoub2022smart}.

We propose \textbf{DeePM}, a \emph{Structured}, \emph{Risk-Robust} \textbf{Dee}p-learning \textbf{P}ortfolio \textbf{M}anager that learns a trading policy end-to-end. Unlike ``black box'' approaches that treat assets as anonymous time series, DeePM injects specific domain structures -- macroeconomic graphs, realizability constraints, and robust risk measures -- directly into the architecture. We train DeePM with a \emph{window-robust} objective that emphasizes the hardest historical subperiods: a differentiable soft-min aggregation of rolling-window Sharpe ratios (defined in Sec.~\ref{sec:objective}).

\begin{figure}[htbp]
    \centering
    \includegraphics[width=\linewidth]{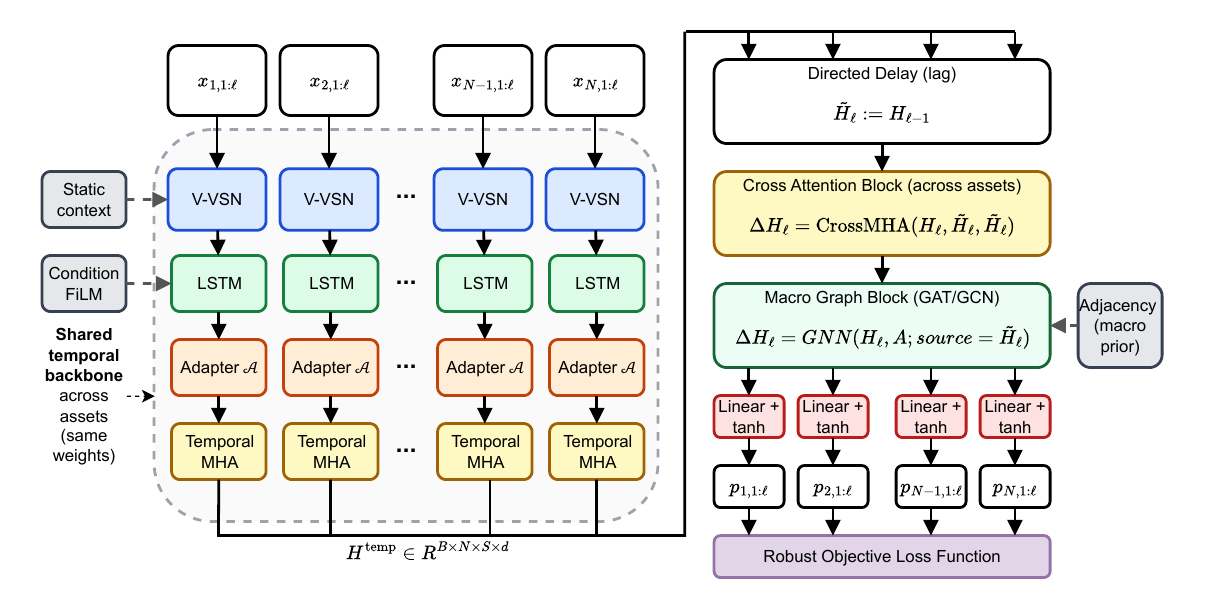}
    \caption{The DeePM pipeline. (1) \textbf{Temporal:} Per-asset history is processed via a hybrid backbone. (2) \textbf{Cross-Sectional:} Assets attend to the global state using a causal Directed Delay. (3) \textbf{Structural/Topological:} Latent embeddings are refined via a Macro-Graph GNN. (4) \textbf{Objective:} The network minimizes a robust loss combining pooled Net Sharpe and a worst-window SoftMin penalty.}
    \label{fig:intro_pipeline}
\end{figure}
\subsection{Decision Problem and Implementation Constraints}\label{sec:intro:decision}
Modern systematic macro trading requires learning sequential portfolio policies from noisy, non-stationary markets while respecting strict realizability constraints, such as asynchronous global closes and path-dependent transaction costs. At each decision time $t$, the agent maps a high-dimensional information set $\mathcal{F}_t$ to a vector of portfolio trading signals or weights $p_t \in \mathbb{R}^{N}$. Unlike stylized frictionless settings, practical deployment faces three binding constraints that define the learning challenge:
\begin{enumerate}
    \item[(i)] \textbf{Asynchronous Information (Ragged Filtration):}
    global macro universes are plagued by \emph{asynchronous market closes} (e.g., Tokyo vs. New York). Naive application of standard attention mechanisms allows models to exploit spurious correlations between ``stale'' and ``fresh'' data, leading to look-ahead bias that vanishes in production \citep{lo1990nonsynchronous}. Concretely, under asynchronous global closes, some assets’ same-day information is not available at the portfolio decision time, even though other markets have already closed. If a cross-asset module mixes \emph{same-day} representations across all assets, it can inadvertently condition on information that would only be known later in the day for some markets, creating a subtle look-ahead that disappears in live trading. Our policy must ensure that any cross-asset features used at time $t$ are measurable with respect to the common information set available across \emph{all} assets at decision time. Furthermore, learning spurious correlations instead of true causal links can lead to backtest overfit \citep{bailey2016pbo}. 
    \item[(ii)] \textbf{Path-Dependent Frictions:} Transaction costs are not merely a post-hoc deduction but a state-dependent penalty on turnover ($|w_t - w_{t-1}|$). To be viable, the policy must learn to internalize execution efficiency, filtering out high-turnover signals that do not survive costs.
    \item[(iii)] \textbf{Regime Fragility:} objective functions based only on pooled averages, such as Sharpe-like criteria \citep{sharpe1966}, can yield seemingly strong performance while concentrating losses into a small number of adverse windows. Risk-measure theory motivates that maximizing average performance is insufficient and we should complement it with tail-sensitive criteria \citep{rockafellar2000cvar,ahmadijavid2012evar}; we require the objective to explicitly penalize the ``worst-case'' windows (minimax), prioritizing survival over average-case maximization.
\end{enumerate}

\subsection{Our Approach: Structured Deep Portfolio Management}
DeePM addresses these challenges via a hierarchical architecture mirroring the workflow of a discretionary macro trader. It processes data in three stages:
\begin{enumerate}
    \item \textbf{Temporal Encoding (Hybrid VSN-LSTM-Attention):} Inspired by the Momentum Transformer \citep{wood2023momentumtransformer_risk}, we employ a specialized per-asset encoder. A \textbf{Vectorized Variable Selection Network (V-VSN)} first filters high-dimensional noisy features. This is followed by an \textbf{LSTM} (Long Short-Term Memory) \citep{hochreiter1997long} backbone to model local path-dependence and denoise volatility, and a \textbf{Temporal Self-Attention} mechanism to capture long-range dependencies and global regime shifts.
    \item \textbf{Cross-Sectional Interaction (The Causal Sieve):} To resolve the asynchrony problem, we employ a \textbf{Directed Delay} protocol ($t-1$) for cross-sectional attention. This transforms the attention mechanism into a differentiable ``Causal Sieve,'' filtering out confounding intraday co-movements to isolate predictive impulse-response signals (Transfer Entropy). As a reference point, we also explore ``cascading'' filtration which maximises information freshness.
    \item \textbf{Structural Regularization (Macro Graph):} A Graph Neural Network (GNN) projects latent representations onto a fixed \textbf{Macroeconomic Prior Graph} (e.g., linking Rates to Foreign Exchange (FX) via carry, or Energy to Inflation). This acts as a spectral low-pass filter, forcing the learned covariance to respect economic topology, which stabilizes performance in low-data regimes.

\end{enumerate}
The entire policy is optimized against a \textbf{Robust Net Sharpe} objective. We introduce a \textbf{SoftMin} penalty on windowed performance, which we show is mathematically equivalent to minimizing the dual form of \emph{Entropic Value-at-Risk (EVaR)} \citep{ahmadijavid2012evar}, effectively training the model against an adversarial reweighting of history (~App.\ref{app:robust:evar}).

The proposed framework offers a unified end-to-end toolkit for practitioners, integrating three distinct inductive biases: \emph{temporal} (learning path-dependent regimes), \emph{cross-sectional} (isolating causal impulse-responses), and \emph{topological} (regularizing via economic structure).

\paragraph{Contributions.}
We make the following contributions:
\begin{enumerate}
    \item \textbf{Decision-focused, cost-aware end-to-end learning.}
    We train portfolio policies directly against realized risk-adjusted net returns, aligning the gradient descent landscape with the investor's utility. This yields net risk-adjusted performance that is approximately double that of two-stage mean–variance baselines and fifty percent higher than that of the Momentum Transformer. The model further learns to naturally dampen turnover without requiring heuristic constraints.
    \item \textbf{The Primacy of Causal Robustness:} Through ablation studies, we show that our conservative \emph{Directed Delay} filtration (using strictly lagged cross-sectional data) outperforms the ``cascading'' filtration that maximizes data freshness. This suggests that in macro trading, identifying robust causal drivers (Transfer Entropy) is more valuable than minimizing information latency.
    \item \textbf{Macroeconomic Graph Regularization:} We show that injecting a sparse prior of economic linkages prevents overfitting. The inclusion of the Graph GNN reduces Maximum Drawdown by \textbf{21\%} relative to a purely data-driven attention model, confirming the value of domain knowledge as an inductive bias.
    \item \textbf{Differentiable EVaR Optimization:} We bridge the gap between deep learning and coherent risk measures by implementing the SoftMin objective. Empirical results show this penalty is the single largest driver of stability, enabling the strategy to maintain positive performance during the post-2020 volatility transition.
    \item \textbf{Optimising Net-of-Cost Returns:} We motivate, theoretically (Sec.~\ref{sec:objective:ensembling}) and empirically, that ensembling to aggregate the trading signals across the top $K$ seeds improves performance net-of-cost and, crucially, training with the full transaction costs leads to suboptimal performance. Our experiments demonstrate that training with a transaction cost scaler of $\gamma=0.5$ improves net performance out of sample by approximately fifty percent, compared to the full cost training.
\end{enumerate}

\subsection{Paper roadmap}\label{sec:intro:roadmap}
Section~\ref{sec:related} reviews related work on portfolio learning under frictions, cross-sectional modelling, structured priors, and robust objectives. Section~\ref{sec:data} formalizes the macroeconomic graph, decision protocol, data construction, and transaction-cost model. Section~\ref{sec:method} introduces DeePM’s architecture, including filtration-respecting cross-asset conditioning and macro graph regularization. Section~\ref{sec:objective} presents the robust learning objective and optimization procedure. Section~\ref{sec:experiments} details the experimental design, reports empirical results, provides ablations and discusses managerial implications. Section~\ref{sec:conclusion} concludes.

\section{Related Literature}\label{sec:related}
We situate DeePM at the intersection of deep time-series momentum, end-to-end portfolio learning, and robust optimization. For a comprehensive study of the techniques we employ in this work, please refer to \citet{zhang2025deep}.

\subsection{Deep Learning for Systematic Trend}\label{sec:related:trend}
Systematic macro strategies typically exploit two primary market phenomena: \emph{Momentum} (or Trend), the tendency for asset prices to persist in their direction over time \citep{jegadeesh1993returns, moskowitz2012tsmom}, and \emph{Mean Reversion}, the tendency for prices to return to a long-term equilibrium after becoming over-extended \citep{debondt1985does}.

Modern systematic trading has moved beyond linear factors toward deep sequence modelling. \citet{lim2019tsmom} introduced \emph{Deep Momentum Networks (DMNs)}, demonstrating that LSTMs could capture non-linear volatility scaling and trend behaviors that linear \emph{Time-series Momentum} (TSMOM) \citep{moskowitz2012tsmom} misses. \citet{wood2023momentumtransformer_risk} extended this with the \emph{Momentum Transformer}, applying multi-head attention to learn regime-switching dynamics and global temporal dependencies. Furthermore, recent advancements \citep{wood2022slowmomentumfastreversion} have integrated online changepoint detection in attempt to adapt to rapid regime shifts. Whilst explicit detection is reactive, a regime-robust approach seeks to internalize these vulnerabilities during training.

While effective, these approaches typically treat assets as independent time series or rely on implicit data-driven correlations. DeePM advances this lineage by explicitly modelling the \emph{cross-sectional} structure of the market. We combine the temporal strengths of the DMN/Transformer architectures with a \emph{Macroeconomic Graph Prior}, regularizing the learning process against known economic linkages to prevent overfitting in low-signal regimes.

\subsection{Machine Learning for Portfolio Construction under Frictions}\label{sec:related:mlpm}
Classical portfolio construction typically estimates parameters $(\mu_t,\Sigma_t)$ and then optimizes a static objective. For mean--variance, one solves
\begin{equation}
\begin{aligned}
w_t^\star
\in \arg\max_{w\in\mathcal{W}}
\left(
\hat{\mu}_t^\top w - \frac{\eta}{2} w^\top \hat\Sigma_t w
\right),
\end{aligned}
\label{eq:mv_classical}
\end{equation}
where $\eta > 0$ represents the investor's risk aversion coefficient. In the unconstrained case, the solution is $w_t^\star=\eta^{-1}\hat\Sigma_t^{-1}\hat{\mu}_t$. A central difficulty is that the optimization step can \emph{amplify} estimation error: because the solution depends on $\hat\Sigma_t^{-1}$, small perturbations in $(\hat{\mu}_t,\hat\Sigma_t)$ can induce large changes in $w_t^\star$. \citet{michaud1989} characterized this as ``error maximization,'' motivating shrinkage, resampling, and risk-based allocations. 

To overcome this, recent works propose \emph{end-to-end} learning. \citet{zhang2020deep} introduced \emph{Deep Learning for Portfolio Optimization}, demonstrating that optimizing the Sharpe ratio directly outperforms two-stage estimation. However, such architectures often rely on stacking asset features. This approach has three limitations: (i) it ignores the natural \emph{permutation symmetry} of the portfolio (the order of assets is arbitrary); (ii) it lacks \emph{structural priors}, forcing the model to learn economic relationships from scratch; and (iii) it often assumes a synchronized data grid, which introduces look-ahead bias in global portfolios with asynchronous closes. DeePM addresses these by using set-invariant attention mechanisms and enforcing a filtration-respecting lag structure.

\subsection{Cross-Sectional Modeling with Attention and Representation Learning}\label{sec:related:attn}
Cross-asset dependence can be modeled with data-dependent interactions over the tradable set $\mathcal{U}_t$. A generic single-head attention layer is
\begin{equation}
\begin{aligned}
h^{(1)}_{i,t}
&= \sum_{j\in\mathcal{U}_t} \alpha_{ij,t}\, V h^{(0)}_{j,t},\\
\alpha_{ij,t}
&= \frac{\exp\!\left(\langle Q h^{(0)}_{i,t}, K h^{(0)}_{j,t}\rangle\right)}
{\sum_{k\in\mathcal{U}_t}\exp\!\left(\langle Q h^{(0)}_{i,t}, K h^{(0)}_{k,t}\rangle\right)}.
\end{aligned}
\label{eq:attn}
\end{equation}
where $h^{(0)}_{i,t}$ denotes the input embedding for asset $i$ at time $t$, $h^{(1)}_{i,t}$ represents the output representation, $\alpha_{ij,t}$ is the attention weight quantifying the influence of asset $j$ on $i$, $Q, K, V$ are the learnable projection matrices, and $\langle \cdot, \cdot \rangle$ signifies the inner product similarity score. To capture heterogeneous market signals across different feature subspaces, this mechanism is expanded to Multi-Head Attention (MHA) by calculating multiple independent attention heads in parallel and concatenating their results into a single enriched representation.

In portfolio problems the cross-section is naturally a \emph{set}: the ordering of assets is arbitrary, and the traded universe can change with missingness, contract rolls, or data availability. A desirable inductive bias is therefore \emph{permutation equivariance}: if we permute the asset order, the output representations (and hence positions) permute in the same way.
Formally, let $\Pi$ be a permutation matrix acting on the asset dimension of the stacked representation $H_t=[h^{(0)}_{1,t},\dots,h^{(0)}_{N,t}]^\top$. Set-based layers satisfy
\begin{equation}
\begin{aligned}
\mathrm{Layer}(\Pi H_t) = \Pi\,\mathrm{Layer}(H_t),
\end{aligned}
\label{eq:perm_equiv}
\end{equation}
which is a standard requirement in set representation learning \citep{zaheer2017deepsets,lee2019settransformer}. Attention of the form \eqref{eq:attn} is permutation equivariant when applied across the asset axis, making it well suited for variable-size universes and existence masking.

Purely permutation-equivariant layers are \emph{anonymous} across elements; in markets, however, assets have persistent identities (e.g., duration-sensitive rates vs.\ energy futures). DeePM incorporates this by injecting a categorical ticker embedding $e_i$ to the per-asset state before cross-sectional interaction. This preserves equivariance in \eqref{eq:perm_equiv} (the permutation acts jointly on $(\tilde{h}^{(0)}_{i,t})_{i\in\mathcal{U}_t}$), while allowing the model to condition on asset-specific effects and long-run heterogeneity.

Permutation-equivariant attention is also a natural mechanism for transferring structure across regimes or assets by matching patterns in a context set. For example, \citet{wood2024fewshot} use cross-attention to adapt trend-following decisions to new regimes from a small number of examples, highlighting the usefulness of attention-based set reasoning in financial time series. Instead of conditioning on a basket of historical sequences for few-shot transfer, we use a cross-sectional context set of contemporaneous assets (or their lag-aligned histories) to transfer information across the universe.

\subsection{Graph Priors and Economic Structure}\label{sec:related:graph}
While data-driven attention can capture arbitrary correlations, financial covariance matrices are notoriously noisy and unstable. Graph Neural Networks (GNNs) offer a mechanism to inject domain knowledge as a structural prior. By defining a sparse adjacency matrix $A$ based on economic fundamentals -- such as supply chains, sector classifications, or macro-correlation regimes -- we impose a relational inductive bias. Critically, In Graph Neural Networks, the graph topology imposes a structural constraint on where information may flow, not what is learned.

The standard Graph Convolutional Network (GCN) \citep{kipf2017gcn} implements a fixed spectral filter:
\begin{equation}
H^{(2)}_t
=
\sigma\!\left(\tilde{A}\, H^{(1)}_t W_1\right),
\label{eq:gcn}
\end{equation}
where $\tilde{A}$ is the normalized adjacency matrix, and $\sigma(\cdot)$ is a non-linear activation function. This operator enforces \emph{isotropic} smoothing -- every neighbor affects the node equally (weighted by degree). In finance, however, economic linkages are time-varying; a ``Safe Haven" link between Bonds and Gold may be strong in crises but weak in growth regimes.

To address this, Graph Attention Networks (GATs) \citep{velickovic2018gat} introduce learnable, anisotropic attention weights $\alpha_{ij}$ masked by the graph structure and a message passing layer then updates node $i$ by aggregating over its graph neighborhood:
\begin{equation}
\begin{aligned}
h_i^{(2)} 
&= \sigma\left( \sum_{j \in \mathcal{N}(i)} \alpha_{ij} W h_j^{(1)} \right), \\
\alpha^{G}_{ij,t}
&= \frac{\exp\!\left(\langle Q_G h^{(1)}_{i,t},\, K_G h^{(1)}_{j,t}\rangle\right)}
{\sum_{k\in\mathcal{N}(i)}\exp\!\left(\langle Q_G h^{(1)}_{i,t},\, K_G h^{(1)}_{k,t}\rangle\right)} ,
\end{aligned}
\label{eq:graph_attn}
\end{equation}
where $\mathcal{N}(i)=\{j : A_{ij}=1\}$ is defined by the (sparse) economic adjacency $A$. Unlike fully connected attention (Section \ref{sec:related:attn}), GAT restricts attention to economically plausible pairs defined by $A$, while unlike GCN, it allows the model to dynamically upweight or downweight these priors based on the current market state. Furthermore, both GCN and GAT layers retain the property of permutation equivariance (under simultaneous permutation of node features and the adjacency matrix), ensuring the model remains consistent with the set-based nature of the portfolio universe discussed in Section \ref{sec:related:attn}. DeePM utilizes this architecture to balance structural regularization with regime-dependent flexibility.

\subsection{Robust Objectives and Risk Measures}\label{sec:related:robust}
Standard portfolio objectives maximize average performance (e.g., pooled Sharpe ratio), which can lead to overfitting on specific regimes or ``lucky windows" while ignoring tail risks. To address this, we draw on the literature of coherent risk measures.
Conditional Value-at-Risk (CVaR) \citep{rockafellar2000cvar} at level $\alpha$ admits the variational form:
\begin{equation}
\mathrm{CVaR}_\alpha(Z)
=
\min_{\eta\in\mathbb{R}}
\left\{
\eta + \frac{1}{1-\alpha}\E[(Z-\eta)_+]
\right\}.
\label{eq:cvar}
\end{equation}
Entropic Value-at-Risk (EVaR) \citep{ahmadijavid2012evar} provides a strictly tighter upper bound via exponential tilting:
\begin{equation}
\mathrm{EVaR}_\alpha(Z)
=
\inf_{\lambda>0}
\frac{1}{\lambda}
\left(
\log \E[\mathrm{e}^{\lambda Z}] - \log \alpha
\right).
\label{eq:evar}
\end{equation}

Crucially, these measures link directly to \emph{distributionally robust optimization} (DRO) or minimax frameworks. By convex duality, minimizing EVaR is equivalent to minimizing the expected loss against an \emph{adversarial distribution} $Q$ chosen from a uncertainty set defined by the Kullback-Leibler (KL) divergence:
\begin{equation}
\mathrm{EVaR}_\alpha(Z)
=
\sup_{Q \ll P} \left\{ \E_Q[Z] \mid \mathrm{KL}(Q \| P) \le C_\alpha \right\},
\label{eq:evar_dual}
\end{equation}
where $C_\alpha=-\log(1-\alpha)$. DeePM's training objective employs a \textbf{SoftMin} penalty, defined in Sec.~\ref{sec:objective:robust_loss}, over windowed Sharpe\footnote{In Sec.~\ref{sec:objective:robust_loss}, we also motivate why we use Sharpe ratio} ratios. As we detail in App.~\ref{app:robust_objective}, this is mathematically isomorphic to the dual form of EVaR. It effectively trains the policy against an implicit adversary who reweights the training windows ($Q$) to emphasize the worst-performing regimes, thereby enforcing robustness against distributional shifts.

\begin{figure}[thb]
    \centering
    \includegraphics[width=0.8\linewidth]{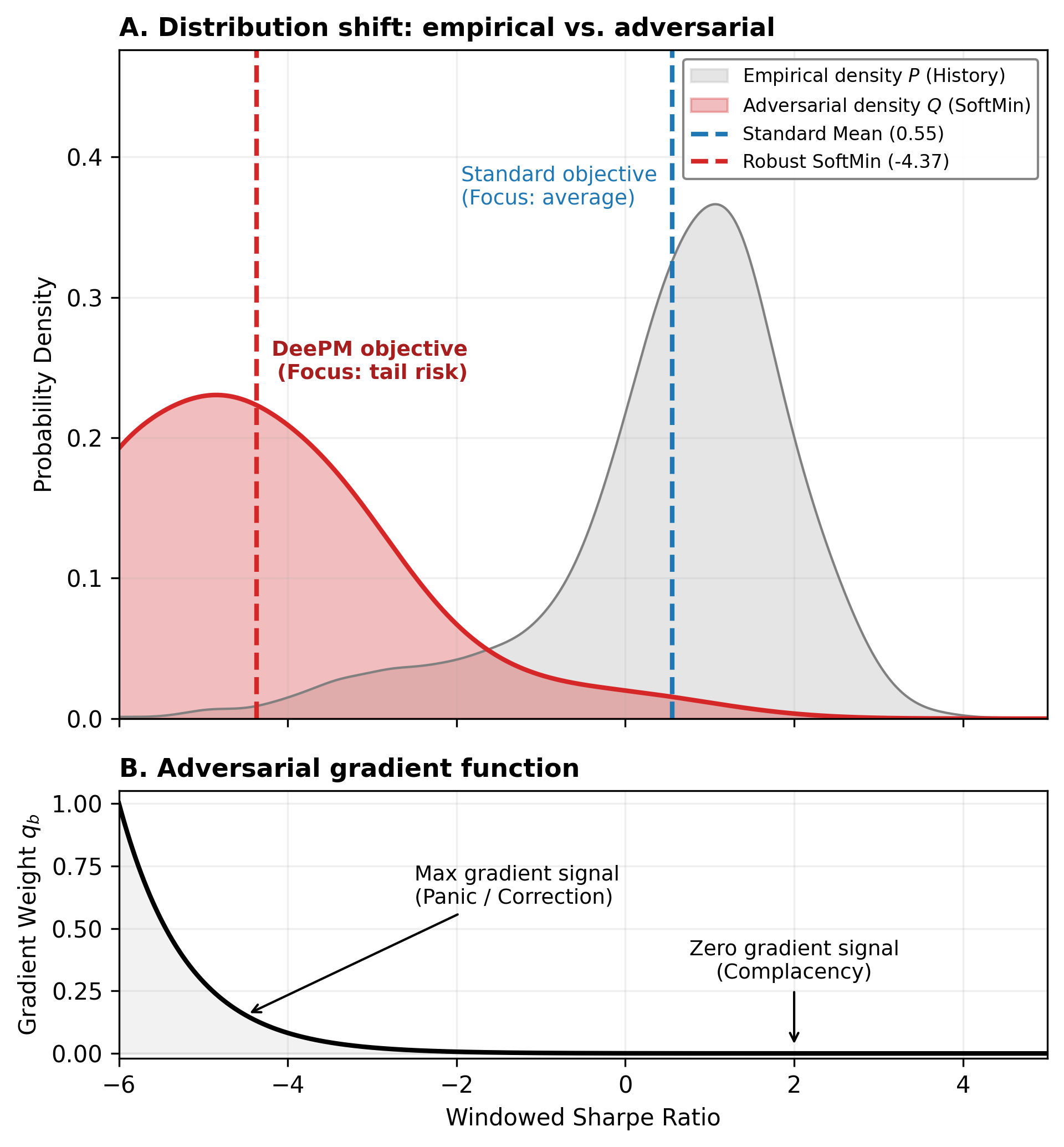}
    \caption{\textbf{Visualizing the Distributionally Robust Objective.} 
    This figure illustrates the mechanics of the SoftMin penalty using a synthetic mixture model (85\% Normal regime $\mathcal{N}(1.0, 0.9^2)$, 15\% Crisis regime $\mathcal{N}(-2.0, 1.4^2)$). 
    \textbf{Panel A} demonstrates the implicit distribution shift: while the empirical history $P$ (grey) has a positive mean ($+0.55$), the adversarial reweighting $Q$ (red) shifts probability mass to the left tail, resulting in a significantly lower robust utility ($-4.37$). This effective ``hallucination'' of a harsher environment forces the optimizer to prioritize survival in worst-case regimes.
    \textbf{Panel B} displays the corresponding gradient weight function $q_b \propto \exp(-\mathrm{SR}_b/\tau)$. The exponential decay ensures that ``easy'' high-Sharpe windows contribute near-zero gradient signal (Complacency), while ``hard'' negative-Sharpe windows dominate the parameter updates (Panic), effectively implementing a differentiable minimax curriculum.}
    \label{fig:robust_objective_viz}
\end{figure}

\section{Problem Setup and Data}\label{sec:data}

\subsection{Universe and Decision Protocol}\label{sec:data:protocol}
We trade a diversified universe of $N=50$ liquid futures and FX contracts. Each asset $i$ is assigned to a macro group $g(i)$ (e.g., Equities, Rates, Energy), which informs the structural graph prior (details in Appendix~\ref{app:universe:graph}).

Systematic macro portfolios face the challenge of \emph{asynchronous market closes} (e.g., Nikkei closes hours before S\&P 500). We evaluate two protocols to define the admissible information set $\mathcal{F}_t$: \textbf{(1) Global One-Day Lag (Primary)} which enforces a strict delay, and \textbf{(2) Cascading Filtration (Ablation)} which maximizes information freshness. These are defined mathematically in Sec.~\ref{sec:method:cross_section}.
Unless otherwise stated, results utilize the Global One-Day Lag to prioritize structural robustness over latency.

We define the next-period arithmetic return as:
\begin{equation}
\begin{aligned}
r_{i,t+1}
\;:=\;
\frac{P_{i,t+1}}{P_{i,t}} - 1.
\end{aligned}
\label{eq:return_def}
\end{equation}
This convention ensures that any cross-asset operation at decision date $t$ uses only
$\{x_{i,t}\}_{i\in\mathcal{U}}$ computed from $\{P_{i,u}\}_{u\le t}$.

\subsection{Portfolio Returns and Frictions}\label{sec:data:returns}
The policy network outputs raw actions $\tilde{a}_{i,t}\in\mathbb{R}$ which are squashed to
bounded risk weights
\begin{equation}
\begin{aligned}
p_{i,t} := \tanh(\tilde{a}_{i,t}) \in (-1,1).
\end{aligned}
\label{eq:tanh_action}
\end{equation}
Let $\hat{\sigma}_{i,t}>0$ be an ex-ante daily volatility estimate (typically a 63-day EWMA). We define the
volatility scaling factor ($\varepsilon$ being a small regularizer)
\begin{equation}
\begin{aligned}
v_{i,t} := \frac{1}{\hat{\sigma}_{i,t}+\varepsilon}.
\end{aligned}
\label{eq:vs_factor}
\end{equation}
The corresponding (vol-targeted) notional exposure \citep{moskowitz2012tsmom, harvey2018impact} is
\begin{equation}
\begin{aligned}
w_{i,t} := v_{i,t}\,p_{i,t}.
\end{aligned}
\label{eq:notional_def}
\end{equation}

We optimize the policy directly against \emph{Net Portfolio Return} ($R^{\text{net}}$). We define $R^{\text{net}}$ by deducting linear transaction costs from the gross volatility-scaled performance:
\begin{equation}
R^{\mathrm{net}}_{t+1} := \underbrace{\frac{1}{N_t}\sum_{i=1}^N m_{i,t}\, p_{i,t}\, y_{i,t+1}}_{R^{\mathrm{gross}}_{t+1}}
 - \underbrace{\frac{\gamma}{N_t}\sum_{i=1}^N m_{i,t}\, c_i \, \bigl|w_{i,t} - w_{i,t-1}\bigr|}_{\text{Transaction Cost}}
\label{eq:net_return}
\end{equation}
where $y_{i,t+1}$ is the vol-scaled asset return, $c_i$ is the asset-specific cost parameter (derived in App.~\ref{app:universe:data}), $m_{i,t}\in\{0,1\}$ is an availability mask indicating whether asset $i$ is tradable/has valid data and $\gamma$ is a global scaling factor. We motivate in App.~\ref{app:turnover_proofs} that an intermediate value $\gamma \in (0,1)$ is theoretically optimal for maximizing out-of-sample net Sharpe under ensembling. This formulation ensures the model explicitly learns to balance signal strength against turnover constraints.

\subsection{Features and Preprocessing}\label{sec:data:features}
We construct a compact, stationarized feature vector $x_{i,t} \in \mathbb{R}^F$ for each asset, designed to capture multi-scale trend dynamics and tail risks. To strictly isolate the contribution of the learned structural priors and attention mechanisms, we restrict the input space to be derived \emph{exclusively} from daily closing prices. While practical systematic macro strategies typically incorporate explicit ``Carry'' signals (e.g., interest rate differentials, forward points) and fundamental macroeconomic indicators (e.g., Consumer Price Index, Purchasing Managers' Index releases), we omit these sources in this study. This constraint forces the model to extract latent risk premia and regime shifts solely from endogenous price dynamics.

\paragraph{Ex-ante Volatility.}
We estimate daily return variance $\hat{s}^2_{i,t}$ via an exponentially weighted moving average (EWMA) with a 63-day span. This serves as the normalization factor for all inputs and notional sizing.

\paragraph{1. Volatility-Normalized Returns.}
To capture momentum across horizons \citep{moskowitz2012tsmom}, we compute returns over windows $h \in \{1, 21, 63, 252\}$ days (with 252 days representing a full year). Each return is scaled by ex-ante volatility to ensure distribution stability:
\begin{equation}
\mathrm{R}^{(h)}_{i,t}
\;:=\;
\frac{P_{i,t} / P_{i,t-h} - 1}{\hat{\sigma}_{i,t}\sqrt{h} + \varepsilon}.
\end{equation}

\paragraph{2. MACD Trend Filters.}
Following \citet{appel1979macd}, we include Moving Average Convergence Divergence (MACD) signals to capture trend persistence. We compute the difference between fast and slow EWMAs at multiple scales $(S, L) \in \{(8,24), (16,48), (32,96)\}$:
\begin{equation}
\mathrm{MACD}_{S,L}(i,t)
\;:=\;
\frac{\text{EWM}_{S}(P_{i})_t - \text{EWM}_{L}(P_{i})_t}{\hat{\sigma}_{i,t}}.
\end{equation}
The resulting MACD signals are subsequently re-normalized by their own 252-day rolling standard deviation to ensure comparability across assets and time.

\paragraph{3. Mean-Reversion}
To detect over-extension and regime fragility \citep{harvey2000conditional,bollinger2002bollinger}, we include
\textbf{Price Z-Scores:} $Z$-score of log-prices over rolling windows $\ell \in \{21, 252\}$ to signal mean-reversion potential.

\paragraph{4. Robust Outlier Control.}
Financial data is heavy-tailed. To prevent gradient explosions, we apply a robust, no-lookahead clipping filter. For each feature $f_{i,t}$, we compute a rolling median ($m_t$) and Median Absolute Deviation ($\text{MAD}_t$) over 252 days. We clip values to $[m_t \pm 5 \times 1.48 \times \mathrm{MAD}_t]$, ensuring the model trains on the core distribution rather than artifacts. We use $\mathrm{MAD}_t$ (rather than the sample standard deviation) because it is far less sensitive to heavy tails and transient price spikes, scale it by $1.48$ to obtain a $\sigma$-comparable robust dispersion estimate under a Gaussian reference model, and, in line with the literature \citep{lim2019tsmom, wood2023momentumtransformer_risk}, use a conservative $5\times$ (approximately ``$5\sigma$'') band so clipping targets only extreme outliers while preserving almost all typical observations.

\paragraph{5. Feature Subsets and Parsimony.}
Given the high capacity of the DeePM architecture, supplying highly correlated inputs (e.g., overlapping return windows simultaneously with MACD filters) increases the risk of noise memorization rather than generalization. To mitigate this, we do not use all features simultaneously. Instead, we evaluate two distinct, parsimonious subsets: (1) a \emph{Raw Momentum} variant, consisting of the lagged returns ($h \in \{1, \dots, 252\}$) and Z-scores; and (2) a \emph{Signal-Based} variant, consisting only of the 1-day return ($r_{1d}$), the MACD trend filters, and Z-scores. This separation forces the model to learn from orthogonal signals and reduces the likelihood of overfitting to redundant feature correlations.

\subsection{Philosophy of the Macro-Structural Prior}\label{sec:data:graph_philosophy}
In high-dimensional macro environments, pure data-driven correlation learning often fails due to low signal-to-noise ratios and non-stationarity. To mitigate this, we inject domain knowledge via a fixed \emph{Macroeconomic Graph Prior}. We model the universe as a graph $\mathcal{G} = (\mathcal{V}, \mathcal{E})$, constructed deterministically from economic first principles rather than noisy return correlations. 
We acknowledge that any hand-specified macro topology entails some subjectivity; accordingly, we treat $\mathcal{G}$ as a \emph{structural regularizer} rather than a claim of ground-truth causality. Our design is (i) \emph{deterministic and ex ante}, specified from broad economic transmission channels (not fit to the sample), and (ii) \emph{coarse and interpretable}, relying only on high-level mechanisms for which there is broad agreement in macro/markets practice. The resulting construction is summarized here and specified in full in App.~\ref{app:universe:graph}.
\begin{figure}[htbp]
    \centering
    \includegraphics[width=0.8\linewidth]{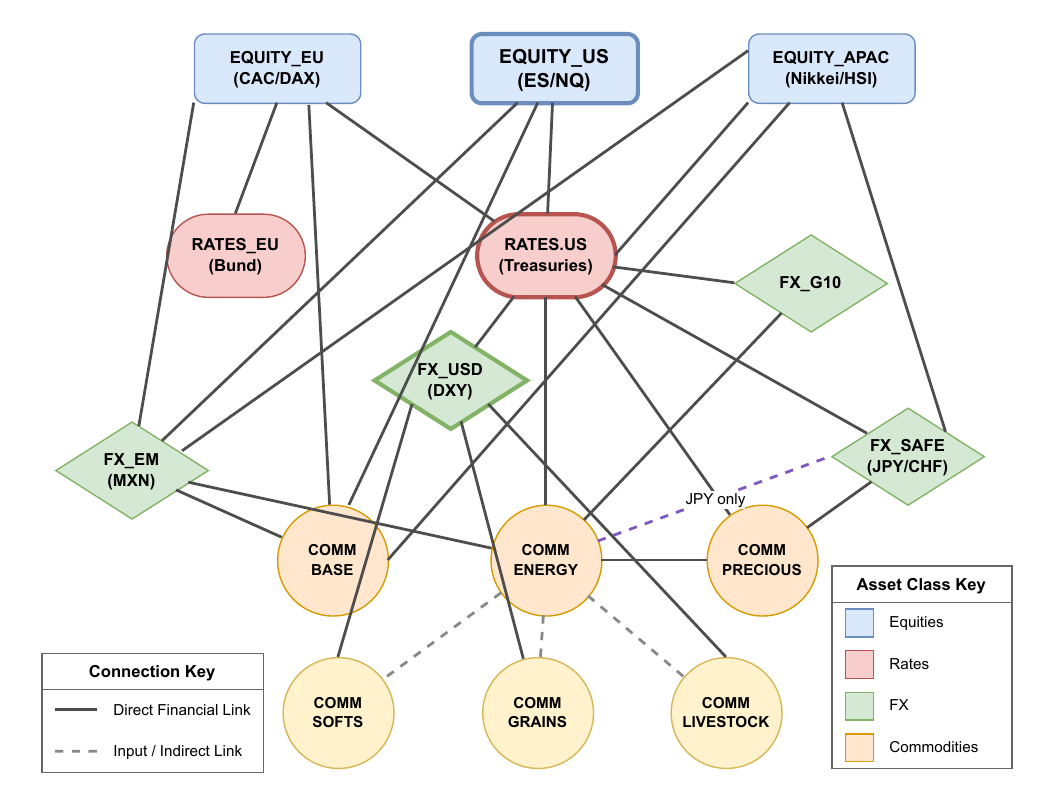}
    \caption{The Macro-Structural Prior Graph used to regularize cross-sectional attention. Edges encode deterministic economic linkages rather than data-driven correlations.}
    \label{fig:macro_graph_prior}
\end{figure}
\begin{enumerate}
    \item \textbf{Sectoral Homophily (Cliques):} Assets within the same macro group (e.g., \texttt{RATES\_US\_TREASURY}) are assumed to share a single latent factor. These form fully connected sub-graphs.
    \item \textbf{Supply Chain \& Substitution:} We encode physical dependencies, such as the input-cost relationship between \textit{Energy} and \textit{Agriculture}, or the substitution effect between \textit{Safe Haven FX} and \textit{Precious Metals}.
    \item \textbf{Macro-Finance Transmission:} We explicitly link disparate asset classes based on canonical transmission mechanisms. For instance, the ``Carry" channel links High-Yield FX to Sovereign Rates, and the ``Inflation Triangle" connects Energy, Breakeven-sensitive Rates, and Precious Metals.
    \item \textbf{Regional Integration:} We capture local monetary and fiscal channels by forming triangular linkages between the primary equity index, sovereign bond, and currency of the same region, enforcing consistency across local asset classes.
\end{enumerate}
This graph serves as a structural scaffold applied \emph{after} the cross-sectional attention mechanism: we require the attention-enriched representations to be smooth with respect to this economic topology, effectively regularizing the high-capacity temporal encoders.

\subsection{Philosophy of the Transaction Cost Model}\label{sec:data:cost_philosophy}
To train a policy that is deployable in practice, the training signal must deduct realistic frictions. A naive constant cost model (e.g., 1bp flat) fails to capture the heterogeneity of global liquidity. Instead, we implement a \emph{Structural Minimum Cost Model} that synthesizes the ``Tick Size Constraint" theory \citep{harris2003trading} with market impact models \citep{kyle1985continuous}.

We define the transaction cost $c_i$ for asset $i$ as a function of its microstructure regime:
\begin{equation}
    c_i \approx C_{\text{struct}, i} \times \lambda_i
\end{equation}
\begin{itemize}
    \item \textbf{The Structural Floor ($C_{\text{struct}}$):} In electronic order books, the quoted spread is lower-bounded by the exchange's Minimum Price Variation (tick size). For liquid instruments, the cost is dominated by this quantization noise.
    \item \textbf{The Liquidity Scalar ($\lambda_i$):} We apply a multiplier $\lambda_i \ge 1.0$ to account for institutional size impact. For deep markets (e.g., S\&P 500), $\lambda \approx 1$. For ``gappy" markets (where there are holes in the order book e.g., Feeder Cattle) or ``Roach Motel" liquidity regimes (easy to enter, hard to exit, e.g., Palladium), $\lambda$ scales significantly higher to reflect the effective spread paid to sweep the book.
\end{itemize}
The exact derivation of these bands and the asset-specific overrides are detailed in App.~\ref{app:universe:tcost}.

\section{Method: DeePM Architecture}\label{sec:method}
The architecture we used is ``structured'', which we use to denote two inductive biases.
First, the cross-sectional module is \emph{permutation equivariant} over assets: relabeling assets only relabels the corresponding outputs,
which is essential when each asset has a categorical identity embedding but the model should not depend on arbitrary ordering.
Second, we incorporate a fixed macro adjacency graph that regularizes cross-asset interactions toward economically plausible transmission channels.
Together these biases reduce overfitting and improve robustness under changing cross-asset correlation regimes.
\subsection{Network Inputs and Context}\label{sec:method:notation}
At each decision time $t$, the model processes a rolling lookback window of length $L$. While the financial return targets and execution logic are defined in Section~\ref{sec:data}, the neural network operates on the following specific tensor representations.

For each asset $i \in \mathcal{U}$, we construct a feature vector $x_{i,t} \in \mathbb{R}^{F}$ (as defined in Section~\ref{sec:data:features}). Let $m_{i,t} \in \{0,1\}$ be the existence mask indicating if asset $i$ is tradable at time $t$. The input to the temporal backbone is the window tensor:
\begin{equation}
X_t := \{x_{i,t-\ell}\}_{i=1,\dots,N;\,\ell=1,\dots,L} \in \mathbb{R}^{N \times L \times F}.
\end{equation}

To allow the shared network weights to learn \emph{asset-specific dynamics} (e.g., distinguishing between mean-reverting FX pairs and trending commodities), we initialize a learnable \emph{Static Asset Embedding} $e_i \in \mathbb{R}^{d}$ for each ticker $i$.
This embedding serves as a persistent identifier that breaks parameter anonymity while maintaining \emph{permutation equivariance}—since the embedding $e_i$ is permuted alongside the asset's dynamic features $x_{i,t}$, the model output is independent of the arbitrary tensor ordering.

The embedding is concatenated with the static transaction cost parameter $c_i$ and projected to form a unified \emph{Static Context Vector} $s_i$:
\begin{equation}
s_i = \text{Linear}([e_i; c_i]) \in \mathbb{R}^{d}.
\end{equation}
As detailed below, this context $s_i$ is injected at two specific locations: (1) to modulate feature selection in the V-VSN, and (2) to initialize the hidden state of the LSTM.

\subsection{Temporal Backbone: Hybrid VSN-LSTM-Attention}\label{sec:method:temporal}
Motivated by state-of-the-art results achieved by \cite{wood2023momentumtransformer_risk} in the univariate setting for systematic macro, we process each asset's time series independently to extract a latent regime embedding $h_{i,t}^{\text{temp}}$. We process using \emph{Channel-independence}, which is a deliberate regularization: by sharing the same temporal encoder weights across assets, the model learns a reusable dynamical feature extractor while avoiding direct cross-channel mixing at the raw-input level, which can encourage fitting to idiosyncratic cross-asset noise \citep{nie2023patchtst}. Cross-asset dependence is instead injected explicitly downstream via the Macro-Graph block (Sec.~\ref{sec:method:graph}). Our backbone integrates three distinct inductive biases: sparsity (VSN), local recurrence (LSTM), and global context (Attention).

\paragraph{1. Vectorized Variable Selection Network (V-VSN).}
\emph{Motivation:} Financial data is high-dimensional and noisy; regime changes often render specific features (e.g., momentum) irrelevant while others (e.g., mean reversion) become dominant.
To address this, we employ a \emph{Vectorized VSN} to dynamically weight input features. Unlike sequential Gated Residual Networks \citep{lim2021temporal}, we implement this efficiently using grouped \emph{1D Convolutions} (Conv1D). Here, a Conv1D with kernel size 1 acts as a time-shared linear projection applied independently to each feature channel across the sequence, enabling efficient parallelization.

We generate feature-wise gating weights via \emph{Feature-wise Linear Modulation (FiLM)} \citep{perez2018film}. Conditioned on the static context $s_i$, FiLM applies a channel-wise affine transformation (scale $\gamma$ and shift $\beta$) to the inputs.
For input $x_{t}$, we compute sparse weights $w_t \in [0,1]^F$ and transformed features $v_t$:
\begin{equation}
\begin{aligned}
\text{FiLM}(x_t, s_i) &= \gamma(s_i) \odot x_t + \beta(s_i), \\
w_t &= \text{Softmax}(\text{Linear}(\text{FiLM}(x_t, s_i))), \\
v_t &= \sum_{f=1}^F w_{t,f} \cdot \text{Conv1D}(x_{t,f}),
\end{aligned}
\end{equation}
where $\odot$ denotes element-wise multiplication. This allows the model to learn, for example, that ``Trend features are less relevant for Asset A (high mean-reversion) than Asset B,'' and suppress them accordingly.

\paragraph{2. Local Recurrence (LSTM).}
\emph{Motivation:} Financial time series exhibit strong path dependence and heteroskedastic noise. Standard Transformers often struggle to filter high-frequency noise without very large datasets.
We pass the weighted feature stream $v_t$ through an LSTM layer. This provides a strong inductive bias for sequential processing and acts as a non-linear low-pass filter, denoising local volatility spikes that can confuse downstream attention mechanisms.
\begin{equation}
h_{i,0}, c_{i,0} = \tanh(W_{\text{0}} s_i), \;\;\; h_{i,t}^{\text{lstm}} = \text{LSTM}(v_{i,t}, h_{i,t-1}^{\text{lstm}}).
\end{equation}
Note that the initial state $(h_{i,0}, c_{i,0})$ is a projection of the asset embedding $s_i$, priming the recurrence with the asset's specific identity.

\paragraph{3. Temporal Attention with Adapter.}
\emph{Motivation:} Regime shifts (e.g., inflation shocks) may depend on events occurring months in the past, beyond the effective memory of an LSTM.
To capture these long-range dependencies, we refine the LSTM state using a causal \emph{Temporal Transformer} \citep{lim2021temporal} block (Temporal MHA).

\emph{The ResSwiGLU Adapter:} Before entering the attention mechanism, the LSTM output is processed by a specialized adapter module $\mathcal{A}(\cdot)$. The adapter is necessary to project the LSTM's robust local features into a richer, non-linear representation that is optimized for the subsequent attention mechanism to capture complex global dependencies. We employ a \emph{Post-Norm} configuration with SwiGLU activation \citep{shazeer2020glu}:
\begin{equation}
\mathcal{A}(x) = \text{LN}\left( x + \delta\bigl( W_2( W_1 x \odot \text{SiLU}(V x) ) \bigr) \right),
\end{equation}
where $W_1, W_2, V$ are learnable weight matrices, $\delta$ denotes \emph{dropout} \citep{srivastava2014dropout}, and $\text{LN}(\cdot)$ is Layer Normalization. Layer-Norm stabilizes training by re-centering and scaling the hidden states to prevent gradient explosion, while Dropout randomly zeroes out neurons during training to force the model to learn robust, redundant features rather than overfitting to financial noise. The residual connection facilitates gradient flow and ensures the model can default to the robust local features extracted by the LSTM if complex non-linear adaptation is unnecessary \citep{he2016resnet}. The term $W_1 x \odot \text{SiLU}(V x)$ represents SwiGLU activation, a non-linear, gated linear unit that improves gradient flow depthwise \citep{shazeer2020glu}, where $\text{SiLU} \colon  z \mapsto \frac{z}{1 + \mathrm{e}^{-z}}$. Unlike Pre-Norm variants \citep{xiong2020layer}, common in modern architectures, the normalization is applied \emph{after} the residual connection, which we empirically found stabilizes the high-variance financial features.

\emph{Note on Architecture Reuse:} This adapter $\mathcal{A}(\cdot)$ serves as the standard non-linear building block throughout DeePM. It is reused as the Feed-Forward Network (FFN) in both the Cross-Sectional Attention layer (Section \ref{sec:method:cross_section}) and the Graph Neural Network layer (Section \ref{sec:method:graph}), ensuring consistent gradient dynamics across temporal and spatial dimensions.

While the LSTM effectively filters high-frequency noise and captures local path dependence, its recursive nature suffers from fading memory over long horizons. To resolve this, the final temporal embedding is produced by applying a causal \emph{Temporal Attention} mechanism. This allows the model to bypass the recursive bottleneck and directly attend to relevant historical regimes (e.g., a past inflation shock similar to the current state) regardless of their temporal distance:
\begin{equation}
 h_{i,t}^{\text{temp}} = \text{TempMHA}(Q=z_{i,t}, K=z_{i,\le t}, V=z_{i,\le t}),
\end{equation}
where $z_{i,t} = \mathcal{A}(h_{i,t}^{\text{lstm}})$ and $\text{TempMHA}(\cdot, \cdot, \cdot)$  is masked to mitigate look-ahead bias. The output is subsequently processed by the adapter $\mathcal{A}(\cdot)$, yielding $H_t = \mathcal{A}(H_t^{\text{temp}})$.

\subsection{Cross-Sectional Interaction: Filtration-Compliant Attention}\label{sec:method:cross_section}
After temporal encoding, we have a tensor of independent asset embeddings $H_t \in \mathbb{R}^{N \times d}$. To capture cross-asset spillover (e.g., rates driving equities), we apply a \emph{Cross-Sectional Multi-Head Attention (MHA)} layer.

\emph{Post-Norm with ReZero Gating:} We implement this block using a Post-Norm configuration augmented with \emph{ReZero} gating \citep{bachlechner2021rezero}. Standard Transformer connections can suffer from signal degradation in deep networks; ReZero mitigates this by introducing a learnable scalar parameter $\alpha$ initialized to a small value (e.g., $\alpha \approx 0$). This initializes the layer as an identity map, allowing the optimizer to gradually introduce cross-sectional interactions only where they provide predictive gain.
\begin{equation}
H_t^{\text{attn}} = \text{LN}\left( H_t + \alpha_{\text{cross}} \cdot \text{CrossMHA}\left(H_t, \tilde{H}_t, \tilde{H}_t\right) \right).
\end{equation}
The output is subsequently processed by the adapter $\mathcal{A}(\cdot)$ defined in Section \ref{sec:method:temporal}, yielding $H_t^{\text{cross}} = \mathcal{A}(H_t^{\text{attn}})$.

To handle the asynchronous nature of global closes rigorously, the Key/Value context $\tilde{H}_t$ is defined via the \emph{Filtration-Compliant Context}:
\begin{enumerate}
    \item \textbf{Directed Delay (Primary Protocol):} We strictly lag the entire cross-section to the previous close, $\tilde{H}_t = H_{t-1}$. This enforces a causal information gap, compelling the model to learn predictive impulse-response functions (Transfer Entropy).
    \item \textbf{Cascading Filtration (Ablation):} We construct a composite context, using close times $t_{\text{c}}$ where asset $i$ sees asset $j$'s state from day $t$ if and only if $i$ closes earlier than $j$ (e.g., US observing Japan).
    \begin{equation}
    \tilde{h}_{j,t}^{(i)} = \mathbb{I}(t_{\text{c}, j} < t_{\text{c}, i}) \, h_{j,t} + \mathbb{I}(t_{\text{c}, j} \ge t_{\text{c}, i}) \, h_{j,t-1}.
    \end{equation}
\end{enumerate}

\begin{figure}[htbp]
    \centering
    \includegraphics[width=\linewidth]{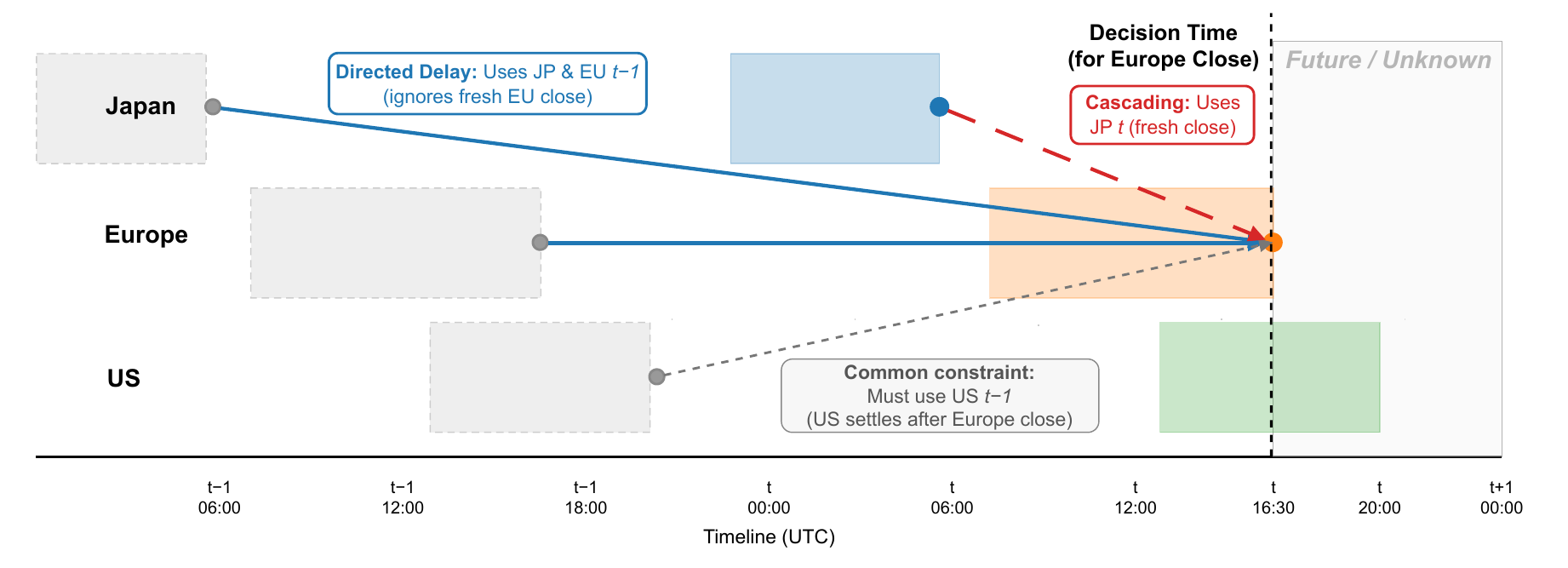}
    \caption{\textbf{The Ragged Filtration Problem.} The timeline illustrates the asynchrony of global closes relative to the portfolio decision time $t$ (Europe Close). 
    A \emph{Cascading Filtration} (red dashed arrow) utilizes the most recent data (in our case we only use close data, but an open could be used), maximizing freshness. 
    DeePM's \emph{Directed Delay} (blue solid arrow) strictly lags cross-sectional attention to $t-1$, enforcing a robust causal gap to isolate predictive impulse-responses.}
    \label{fig:ragged_filtration}
\end{figure}

\subsection{Structural Regularization: Macro Graph Prior}\label{sec:method:graph}
Data-driven attention can overfit in low-signal regimes. We regularize the cross-sectional representations using a \emph{Graph Attention Network (GAT)} constrained to the fixed \emph{Macroeconomic Prior Graph} $\mathcal{G}=(\mathcal{V}, \mathcal{E})$.

\paragraph{Structure and Edge Biases.}
Similar to the cross-sectional layer, we use a \emph{Post-Norm ReZero} configuration. However, unlike standard GATs which often ignore edge strength, we strictly enforce the economic prior by injecting the adjacency weights directly into the attention mechanism. Inspired by the \emph{Graphormer} framework \citep{ying2021graphormer}, instead of using a hard mask based on edges, we add a \emph{fixed structural bias} derived from the adjacency matrix $A$ to the attention scores:
\begin{equation}
\alpha_{ij,t} = \text{Softmax}_j \left( \frac{(Q h_{i,t})^\top (K \tilde{h}_{j,t})}{\sqrt{d}} + \ln(A_{ij} \right),
\end{equation}
where we handle $A_{ij}=0$ by masking non-edges, setting the corresponding logit to $-\infty$. Since $A_{ij}$ encodes the strength of the economic linkage (or $0$ for no link), the term $\ln(A_{ij})$ acts as a soft-masking mechanism: strong links ($A_{ij} \approx 1$) pass the data-driven attention signal unhindered, while weaker or non-existent links ($A_{ij} \to 0$) introduce large negative penalties, effectively pruning the connection.

The GNN aggregation serves as a residual update to the embeddings. The update is as follows:
\begin{equation}
\begin{aligned}
h_{i,t}^{\text{gnn}} &= \text{LN}\left( h_{i,t}^{\text{cross}} + \alpha_{\text{gnn}} \cdot \sum_{j \in \mathcal{N}(i)} \alpha_{ij,t} W \tilde{h}_{j,t} \right), \\
h_{i,t}^{\text{final}} &= \mathcal{A}(h_{i,t}^{\text{gnn}}),
\end{aligned}
\end{equation}
where $\mathcal{N}(i)$ are neighbors defined by the macro graph and $\alpha_{\text{gnn}}$ is the ReZero gate.
To rigorously validate the necessity of this anisotropic attention mechanism, we also implement an \emph{Isotropic GCN} ablation \citep{kipf2017gcn}. In the GCN variant, the learnable attention weights $\alpha_{ij,t}$ are replaced by fixed spectral weights $1/\sqrt{\text{deg}(i)\text{deg}(j)}$, forcing the model to aggregate all economic neighbors equally regardless of their current predictive relevance.

\subsection{Theoretical Interpretations}\label{sec:method:theory}
The architectural choices in DeePM can be unified under two theoretical frameworks: dynamical systems reconstruction and approximate Bayesian inference.

\paragraph{Temporal Encoder as State Space Reconstruction.}
Financial markets can be viewed as high-dimensional, non-stationary dynamical processes observed through
noisy, partial measurements. Motivated by \emph{Takens' Embedding Theorem}
\citep{takens1981detecting, sauer1991embedology}, delay-coordinate histories can provide a sufficient
representation of latent state under suitable regularity conditions. Our temporal backbone
(V-VSN $\to$ LSTM $\to$ TempMHA) therefore learns a data-driven delay map
$\Phi(x_{t-L:t}) \mapsto h_t$, projecting the observable price/return history into a latent state
space in which predictive structure is easier to model. Using volatility-normalized returns helps
stabilize scale over time and across assets (reducing heteroskedasticity), which empirically improves
the conditioning of this learned representation.

\paragraph{Graph Layer as a Bayesian Structural Prior.}
Standard attention mechanisms approximate a purely data-driven posterior $p(Z \mid \mathcal{D})$. In finance, where the signal-to-noise ratio is low, this often leads to overfitting on spurious correlations. We rigorously interpret the Macro-Graph component as injecting a \emph{Gaussian Markov Random Field (GMRF) Prior} over the asset embeddings, $p(Z \mid \mathcal{S}) \propto \exp(-\frac{\alpha}{2} \operatorname{Tr}(Z^\top \mathcal{L} Z))$, where $\mathcal{L} = I - D^{-1/2}AD^{-1/2}$ is the normalized graph Laplacian \citep{rue2005gmrf}. The trace term represents the \emph{Dirichlet energy} of the signal on the graph \citep{chung1997spectral}, which expands as:
\begin{equation}
\operatorname{Tr}(Z^\top \mathcal{L} Z) = \frac{1}{2} \sum_{i,j} A_{ij} \left\| \frac{z_i}{\sqrt{d_i}} - \frac{z_j}{\sqrt{d_j}} \right\|_2^2.
\end{equation}
Minimizing this energy forces economically linked assets (e.g., crude oil and energy equities) to have similar latent representations, scaled by their connectivity, unless the data strongly suggests otherwise. Consequently, the architecture approximates the Maximum A Posteriori (MAP) estimate:
\begin{equation}
\hat{Z}_t \approx \arg\max_Z \Bigl\{ \underbrace{\log p(\mathcal{D} \mid Z)}_{\mathclap{\text{Likelihood (Attn)}}} - \underbrace{\tfrac{\alpha}{2} \operatorname{Tr}(Z^\top \mathcal{L} Z)}_{\mathclap{\text{   Struct. Prior (Graph)}}} \Bigr\}.
\end{equation}
Both the GCN and GAT layers can be viewed as implementing a gradient step on this objective. In high-noise regimes, the prior dominates, smoothing representations across the economic graph to prevent overfitting. In high-signal regimes, the attention mechanism (Likelihood) can override the prior to capture novel market dynamics that deviate from historical economic structures. While a GCN uses the fixed isotropic prior $\mathcal{L}$, the GAT layer allows for an anisotropic refinement where the model learns to dynamically adjust the strength of specific economic linkages ($A_{ij}$) based on the current market regime.

\paragraph{Directed Delay as Information-Theoretic Filtering.}
The choice of filtration protocol fundamentally alters the statistical dependency learned by the attention mechanism. A \emph{Synchronous} approach ($H_t \to H_t$) effectively learns \emph{Mutual Information} $I(X_t; Y_t)$, which is symmetric and dominated by instantaneous co-movements. While high mutual information implies strong correlation, it provides no insight into the direction of influence, making it brittle to regime shifts where correlations break.

By enforcing the strict \emph{Directed Delay} ($H_{t-1} \to H_t$), we constrain the Cross-Sectional Attention to approximate \emph{Transfer Entropy} \citep{schreiber2000transfer}, a non-linear generalization of Granger Causality\footnote{In the linear-Gaussian VAR setting, transfer entropy is equivalent (up to log-base/scaling constants) to Granger causality \citep{barnett2009granger}.}
 \citep{granger1969}. Transfer Entropy $\mathcal{T}_{j \to i}$ measures the reduction in uncertainty about asset $i$'s future state given asset $j$'s past, \emph{conditional} on asset $i$'s own history:
\begin{equation}
\mathcal{T}_{j \to i} = H(x_{i,t} \mid x_{i,t-1}) - H(x_{i,t} \mid x_{i,t-1}, x_{j,t-1}).
\end{equation}
While our model optimizes a Sharpe utility rather than explicitly calculating Shannon entropy, the objectives are aligned: maximizing the Sharpe ratio requires minimizing the variance of the residual return. Under standard Gaussian assumptions\footnote{Whilst raw returns are clearly not Gaussian, we volatility-target returns to better approximate homoskedasticity; outside exogenous shocks and abrupt regime shifts, an approximately Gaussian model is a reasonable working assumption.}, minimizing variance is equivalent to minimizing differential entropy. Consequently, a high attention weight $\alpha_{ij}$ implies that the latent state of asset $j$ at $t-1$ resolves uncertainty about asset $i$ at $t$ (predictive utility) that was not resolved by $i$'s own temporal backbone. This asymmetry enables the model to learn structural market hierarchies (e.g., US Rates driving Emerging Market FX) rather than just phenomenological clusters. In our ablation studies (Section \ref{sec:results}), this ``Causal Sieve'' property proves more valuable than the data freshness offered by the Cascading Filtration, suggesting that out-of-sample robustness relies on identifying \emph{drivers} rather than just \emph{peers}.

\section{Learning Objective and Optimization}\label{sec:objective}
Our training process optimizes the realized \emph{Net Portfolio Return} stream $\mathcal{R} = \{R^{\mathrm{net}}_{b,t}(\theta) \mid b=1\dots B, t=1\dots T\}$, representing the returns across $B$ independent sequences (batches) over a horizon of $L$ time-steps. Unlike standard supervised learning, which minimizes MSE, we maximize a differentiable utility function targeting robust risk-adjusted performance. Return premia are regime-dependent and can unwind abruptly (e.g., momentum “crashes” and fast reversals), motivating objectives that do not concentrate risk in a small number of adverse windows. \citep{wood2022slowmomentumfastreversion,daniel2016momentumcrashes}. 

We specifically focus on the \emph{Sharpe Ratio} \citep{sharpe1966} rather than raw returns for two reasons:
\begin{enumerate}
    \item \textbf{Risk Adjustment:} Maximizing the Sharpe ratio aligns the training objective with the mandate of a portfolio manager, who seeks to maximize returns per unit of risk rather than absolute performance.
    \item \textbf{Statistical Significance:} Maximizing the Sharpe Ratio ($\mathrm{SR}$) is mathematically equivalent to maximizing the $t$-statistic of the strategy's expected return\footnote{This equivalence holds under i.i.d. elliptical returns with finite variance.}. For a batch of size $BL$, the $t$-statistic testing the null hypothesis of zero mean return ($H_0: \mu=0$) is given by:
    \begin{equation}
    t = \frac{\hat{\mu}}{\hat{\sigma} / \sqrt{BL}} = \sqrt{BL} \cdot \frac{\hat{\mu}}{\hat{\sigma}} = \sqrt{BL} \cdot \mathrm{SR}.
    \end{equation}
    Since the window length $BL$ is constant during a training step, the gradient $\nabla_\theta t$ is proportional to $\nabla_\theta \mathrm{SR}$. This aligns the optimization landscape with the objective of finding statistically robust alpha, explicitly penalizing high-variance strategies that might otherwise appear profitable due to ``lucky" outliers.
\end{enumerate}

\subsection{Robust Objective: Pooled Sharpe with SoftMin Penalty}\label{sec:objective:robust_loss}
Standard Sharpe maximization often yields policies that ``cheat" by overfitting to specific high-return windows while ignoring latent tail risks. To mitigate this, we partition the training sequence into $B$ non-overlapping windows (e.g., quarterly blocks) and optimize a hybrid objective:
\begin{equation}
\mathcal{L}(\theta)
=
-\underbrace{\mathrm{SR}_{\mathrm{pool}}(\mathcal{R})}_{\text{Long-Run}}
-
\lambda \cdot \underbrace{\mathrm{SoftMin}_\tau\Big(\{\mathrm{SR}_{b}\}_{b=1}^B\Big)}_{\text{Robustness}}.
\end{equation}

\paragraph{1. Pooled Sharpe Ratio.}
$\mathrm{SR}_{\mathrm{pool}}$ treats the entire mini-batch history (spanning $B$ sequences of length $L$) as a single contiguous equity curve. It explicitly targets the global risk-adjusted return:
\begin{equation}
\mathrm{SR}_{\mathrm{pool}}(\mathcal{R}) := \frac{\hat{\mathbb{E}}[\mathcal{R}]}{\sqrt{\widehat{\mathrm{Var}}[\mathcal{R}] + \varepsilon}}
\end{equation}
where $\hat{\mathbb{E}}[\cdot]$ and $\widehat{\mathrm{Var}}[\cdot]$ denote the empirical mean and variance operators over the full set of realized net returns across all $B$ blocks and $L$ time-steps. Specifically, the pooled sample mean is $\hat{\mathbb{E}}[\mathcal{R}] := \frac{1}{BL}\sum_{b=1}^B\sum_{t=1}^L R^{\mathrm{net}}_{b,t}$, with $\varepsilon$ serving as a small numerical stability constant.

\textbf{Mathematical Motivation (Consistency and Bias):}
Assume the strategy return process is stationary and ergodic.\footnote{Financial returns are not strictly stationary over long horizons due to structural breaks and evolving market microstructure; we adopt stationarity/ergodicity as a useful idealization that justifies interpreting empirical moments as long-run expectations.}
The pooled metric $\mathrm{SR}_{\mathrm{pool}}$ uses an effective sample size on the order of $N_{\text{eff}} \approx B\times L$ (up to dependence adjustments). Under ergodicity, the sample mean and variance converge to their population counterparts, and by the continuous mapping theorem $\mathrm{SR}_{\mathrm{pool}}$ converges (in probability) to the population Sharpe ratio as $BL \to \infty$.
In contrast, the naïve average of window-wise Sharpes,
$\overline{\mathrm{SR}}=\frac{1}{B}\sum_{b=1}^B \widehat{\mathrm{SR}}_b$,
aggregates \emph{ratios} computed on shorter samples and is therefore generally biased and higher-variance in finite samples (since $\widehat{\mathrm{SR}}_b$ is itself a biased ratio estimator).
Moreover, following \citet{lo2002statistics}, the variance of $\widehat{\mathrm{SR}}$ scales as $O(1/L)$ (under weak dependence); pooling increases the effective sample size, reducing gradient noise and stabilizing optimization.

\paragraph{2. SoftMin as an Adversarial Prior.}
To ensure the policy survives adverse regimes, we penalize the smooth minimum of the per-window Sharpe ratios $\{\mathrm{SR}_b\}$. The SoftMin function with temperature $\tau$ is defined as:
\begin{equation}
\mathrm{SoftMin}_\tau(\{\mathrm{SR}_b\}) := -\tau \log \left( \frac{1}{B} \sum_{b=1}^B \mathrm{e}^{-\frac{\mathrm{SR}_b}{\tau}} \right).
\end{equation}
As discussed in Section~\ref{sec:related:robust}, minimizing $-\mathrm{SoftMin}$ is mathematically equivalent to minimizing the dual form of \emph{Entropic Value-at-Risk (EVaR)} \citep{ahmadijavid2012evar}. This effectively trains the model against an implicit adversary who reweights the training windows to emphasize the worst-performing periods (minimax optimization).

The temperature parameter $\tau > 0$ governs the aggressiveness of this adversary. As $\tau \to 0$, the function approaches the hard minimum ($\min_b \mathrm{SR}_b$), forcing the model to focus exclusively on the single worst-performing window (pure minimax). Conversely, as $\tau \to \infty$, the function converges to the arithmetic mean, recovering a standard risk-neutral objective. In practice, $\tau$ acts as a tunable ``robustness control": lower values produce more conservative policies that prioritize survival in difficult regimes, while higher values allow the model to focus on average-case performance.

Empirically, we find that this soft adversarial mechanism significantly stabilizes training. By penalizing potential collapse in ``hard" windows, the gradient descent is prevented from greedily overfitting to ``easy" low-volatility regimes, a common pathology in financial deep-learning. For an in-depth analysis, including the connection to EVaR, see App.~\ref{app:robust_objective}.

\subsection{Ensembling and Implicit Cost Regularization}\label{sec:objective:ensembling}
Deep reinforcement learning is notoriously sensitive to initialization. We mitigate this via an ensemble of the top $K$ models, a practice well-supported in the literature for predictive uncertainty estimation \citep{lakshminarayanan2017deepensembles}. Beyond variance reduction, we prove that ensembling acts as a structural regularizer for transaction costs.

\paragraph{Proposition 1 (Convexity of Turnover Cost).}
Let $\mathcal{C}(p) = \gamma \sum_t |\Delta w_t(p)|$ be the proportional transaction cost function associated with a policy $p$. This function is convex in $p$.

\paragraph{Corollary (Jensen's Inequality for Trading).}
For an ensemble of $K$ independent policies $\{p^{(k)}\}_{k=1}^K$ and their average $\bar{p} = \frac{1}{K}\sum p^{(k)}$, Jensen's inequality implies:
\begin{equation}
\mathcal{C}(\bar{p}) \;\le\; \frac{1}{K}\sum_{k=1}^K \mathcal{C}(p^{(k)}).
\end{equation}
\textbf{Implication:} The trading cost of the ensemble is strictly upper-bounded by the average cost of its constituents. Idiosyncratic ``noise trades" made by individual models tend to cancel out in the average, structurally reducing turnover and improving the net Sharpe ratio.

This theoretical guarantee informs our hyperparameter choice for the explicit turnover penalty $\gamma$ (Eq. \ref{eq:net_return}). As shown in Appendix~\ref{app:turnover_proofs}, the optimal training penalty $\gamma^\star$ is strictly lower for ensembles than for single models. We essentially ``outsource" part of the regularization burden to the ensemble mechanism, allowing individual models to remain responsive to valid signals.

\subsection{Optimization Protocol}\label{sec:objective:train}
We train using AdamW with a rolling window approach. To ensure the Sharpe ratio statistics in Eq. (1) are stable, we require a large effective batch size (e.g., spanning multiple years of data).
Since GPU memory is limited, we introduce \emph{Exact Gradient Accumulation} for global statistics. We accumulate the sufficient statistics ($\sum R_t, \sum R_t^2$) across micro-batches to compute the exact $\mu$ and $\sigma$ for the full logical batch before performing the backward pass, detailed in Sec.~\ref{app:analyatical_sharpe_grads}. This ensures the optimization landscape is invariant to GPU memory constraints.

Financial time series require a warm-up period for recursive states (e.g., LSTM hidden states) to stabilize. We implement a \emph{burn-in} of $L_0=21$ steps (approx. 1 month). During this phase, the model processes inputs to update its internal state, but gradients are masked and returns do not contribute to the loss function. This ensures that only ``mature" representations, fully conditioned on the valid history, are passed to the optimizer, preventing initial state noise from corrupting the learning process.

\section{Experimental Design}\label{sec:experiments}

\subsection{Benchmarks}\label{sec:experiments:benchmarks}
To isolate the contributions of DeePM's end-to-end architecture, we compare against three benchmark families: (A) passive/rule-based allocations, (B) classical trend-following signals (producing risk weights directly), and (C) two-stage ``signal $\rightarrow$ allocation'' pipelines that combine a forecast signal with a covariance-aware risk allocator.

All benchmarks are evaluated under the same execution and return accounting as DeePM. Risk weights $p_{i,t}$ map to vol-targeted notionals via Eq.~\eqref{eq:notional_def}, and net performance is computed using the same transaction-cost model. Throughout, the covariance input $\hat{\Sigma}_t$ to all two-stage allocators is estimated on vol-scaled daily returns using a 252-day rolling window with Ledoit--Wolf shrinkage \citep{ledoit2004wellconditioned}.

\subsubsection{Passive and Rule-Based}
\begin{itemize}
    \item \textbf{Passive Equal Risk:} A constant long risk weight $p_{i,t} \equiv 1$ for all tradable assets. Under the volatility targeting framework (Eq.~\eqref{eq:notional_def}), this ensures that $w_{i,t} \propto 1/\hat{\sigma}_{i,t}$, representing a static risk-parity exposure to global macro risk premia without active timing.
\end{itemize}

\subsubsection{Classical Trend-Following}
These strategies generate risk weights $p_{i,t}$ directly from univariate price history.
\begin{itemize}
    \item \textbf{TSMOM (Time-Series Momentum):} Following \citet{moskowitz2012tsmom}, positions are determined by the sign of the past 12-month return:
    \begin{equation}
        p^{\mathrm{TS}}_{i,t} = \mathrm{sign}\left(r_{i,t-252:t}\right) = \mathrm{sign}\left(\frac{P_{i,t}}{P_{i,t-252}} - 1\right).
    \end{equation}
    
    \item \textbf{Multi-Scale MACD:} We implement a continuous trend signal using volatility-normalized Moving Average Convergence Divergence (MACD) indicators across three time-scales $(S, L) \in \{(8,24), (16,48), (32,96)\}$. The raw MACD values are mapped to positions via a sigmoidal response function $\phi(x) = x \mathrm{e}^{-x^2/4} / 0.89$ \citep{baz2015dissecting, lim2019tsmom} to squash outliers while maintaining linearity near zero:
    \begin{equation}
        p^{\mathrm{MACD}}_{i,t} = \frac{1}{3}\sum_{k=1}^3 \phi\left(\frac{\text{EWM}_{S_k}(P_i) - \text{EWM}_{L_k}(P_i)}{\hat{\sigma}_{i,t}}\right).
    \end{equation}
\end{itemize}

\subsubsection{Two-Stage Signal--Allocation Baselines}
These baselines separate forecasting from portfolio construction. First, a signal generator produces a raw cross-sectional vector $s_t \in \mathbb{R}^N$ (representing the TSMOM or MACD values across all $N$ assets at time $t$). Second, a covariance-aware allocator maps this vector $s_t$ to final risk weights $p_t$.

\begin{itemize}
    \item \textbf{Risk Managed Trend:} A scalar approach where raw trend signals $s_t$ (from TSMOM or MACD) are normalized to unit leverage $\tilde{p}_t = s_t / \|s_t\|_1$ and then scaled to target portfolio volatility $\sigma_{\text{tgt}}$ using the ex-ante covariance matrix $\hat{\Sigma}_t$:
    \begin{equation}
        p_t = \frac{\sigma_{\text{tgt}}}{\sqrt{\tilde{p}_t^\top \hat{\Sigma}_t \tilde{p}_t}} \tilde{p}_t.
    \end{equation}

   \item \textbf{Rolling MVO (Mean--Variance Optimization):} Following the classical mean--variance form in \eqref{eq:mv_classical}, we proxy expected returns by the trend signal ($\hat{\mu}_t \propto s_t$) and stabilize the inversion with ridge regularization. The resulting positions are
    \begin{equation}
    p_t^{\text{MVO}} \;\propto\; \bigl(\hat{\Sigma}_t + \lambda I\bigr)^{-1} s_t,
    \label{eq:mvo_baseline}
    \end{equation}
    where $\hat{\Sigma}_t$ is a rolling shrinkage covariance estimate and $\lambda>0$ controls conditioning. The covariance $\hat{\Sigma}_t$ is estimated using Ledoit-Wolf shrinkage \citep{ledoit2004wellconditioned} over a rolling 252-day window.
    To manage turnover, we optionally add a quadratic turnover penalty $\kappa \|\mathbf{p}_t - \mathbf{p}_{t-1}\|_2^2$, yielding an analytic update that anchors the portfolio to the previous day's weights:
    \begin{equation}
    p_t^{\text{MVO-TP}} \propto (\hat{\Sigma}_t + \kappa I)^{-1} (s_t + \kappa p_{t-1}).
    \end{equation}
    For the TSMOM two-stage pipeline, we set $\kappa=10$. Given that our assets are scaled to unit daily variance ($\sigma^2 \approx 1$) and the input signals are binary ($\pm 1$), a high penalty is required to dominate the covariance term. This dampens the response to instantaneous signal flips, creating a slow-moving anchor that naturally suppresses noise and limits leverage without requiring a hard cap.

    \item \textbf{Risk Parity (ERC):} An Equal Risk Contribution allocator \citep{maillard2010erc}. We solve for long-only weights $q_t \in \mathbb{R}^N_{\ge 0}$ such that every asset contributes equally to ex-ante portfolio risk. Formally, the risk contribution of asset $i$ is defined as $\text{RC}_i = q_i \frac{(\hat{\Sigma}_t q_t)_i}{\sqrt{q_t^\top \hat{\Sigma}_t q_t}}$. We optimize $q_t$ such that $\text{RC}_i = \text{RC}_j \forall i,j$. Finally, we impose the directionality of the original signal vector $s_t$ via element-wise multiplication:
    \begin{equation}
        p_t = \text{sign}(s_t) \odot q_t.
    \end{equation}
    This ensures that while the position \textit{magnitudes} are determined by the optimizer to equalize risk, the \textit{direction} (long or short) is strictly preserved from the trend signal.
\end{itemize}

\subsubsection{Learning-Based Baselines}
\begin{itemize}
    \item \textbf{Momentum Transformer:} A variant of the architecture proposed by \citet{wood2023momentumtransformer_risk}, which corresponds to the temporal encoder of DeePM. This serves as our primary deep learning baseline to quantify the marginal value of DeePM's Graph and Directed Delay components (see \citet{wood2023momentumtransformer_risk} for comparisons of the Momentum Transformer against other simpler models). It slightly deviates from the original \emph{Momentum Transformer} by updating components such as the adapter blocks and VSN to align with more recent developments. Additionally, for fairness, we use the same transaction cost regularizer as DeePM. 
\end{itemize}

\subsection{Ablation Study Design}\label{sec:experiments:ablations}
To rigorously evaluate the individual contributions of DeePM's architectural components and training objectives, we conduct a systematic ablation study. Our experimental design is structured to test four core hypotheses: (1) that the explicit macroeconomic graph structure acts as a necessary regularizer for cross-asset learning; (2) that the directed delay mechanism is crucial for learning causal rather than spurious correlations; (3) that the robust SoftMin objective provides superior stability; (4) that end-to-end internalization of transaction costs outperforms heuristic post-processing; and (5) that end-to-end internalization of transaction costs is most effective when trained with a relaxed scalar $\gamma < 1$, as per Eqn.~\eqref{eq:net_return}, (defaulting to $\gamma=0.5$ in our standard configuration) to balance turnover penalization against signal capture. By systematically removing or replacing these components from the best-performing ensemble configuration, we aim to isolate their marginal value to the final risk-adjusted performance.

\subsection{Evaluation Metrics}\label{sec:experiments:metrics}
For fair comparison across strategies with different intrinsic risk profiles, all out-of-sample return series are rescaled \emph{ex-post} to a uniform annualized volatility target of $\sigma_{\text{tgt}} = 10\%$. Let $R_t$ denote the rescaled daily net return and $P_t$ the cumulative wealth index at time $t$.

\paragraph{Performance and Risk.}
\begin{itemize}
    \item \textbf{Net Sharpe Ratio (SR):} The primary metric for risk-adjusted return efficiency after transaction costs \citep{sharpe1966}. Defined as the annualized mean return divided by the annualized volatility:
    \begin{equation}
        \text{SR} = \sqrt{252} \cdot \frac{\widehat{\mathbb{E}}[R_t]}{\sqrt{\widehat{\text{Var}}[R_t]}}.
    \end{equation}
    We also report \textbf{Gross SR} to quantify the ``execution gap" (performance degradation due to trading frictions). This metric tests whether the strategy generates excess returns per unit of total risk.
    
    \item \textbf{Compound Annual Growth Rate (CAGR):} The geometric rate of return, capturing the effect of volatility drag on long-term wealth accumulation:
    \begin{equation}
        \text{CAGR} = \left( \prod_{t=1}^T (1 + R_t) \right)^{252/T} - 1.
    \end{equation}
    This metric evaluates the final wealth multiplier available to an investor, accounting for the compounding of losses.

    \item \textbf{Maximum Drawdown (MDD):} The largest peak-to-trough decline in the cumulative equity curve over the test period \citep{magdon2004mdd}, measuring the worst historical loss:
    \begin{equation}
        \text{MDD} = \min_{t \in [0,T]} \left( \frac{P_t}{\max_{s \le t} P_s} - 1 \right).
    \end{equation}
    This tests the strategy's tail risk and potential for capital preservation during adverse regimes.

    \item \textbf{Calmar Ratio:} A tail-risk-adjusted performance measure defined as the ratio of annualized return to absolute maximum drawdown \citep{young1991calmar}:
    \begin{equation}
        \text{Calmar} = \frac{\text{CAGR}}{|\text{MDD}|}.
    \end{equation}
    This metric assesses whether returns are sufficient to compensate for the psychological and financial cost of deep drawdowns.

    \item \textbf{Heteroskedasticity and Autocorrelation Consistent (HAC) $t$-statistic ($t$):} The $t$-statistic for the null hypothesis that the strategy's mean net return is zero ($H_0: \mu = 0$). To ensure valid inference under the serial correlation and heteroskedasticity typical of financial time series, we utilize Newey-West standard errors \citep{neweywest1987}. This provides a statistical significance test for the existence of a positive risk premium.
\end{itemize}

\paragraph{Execution and Turnover.}
\begin{itemize}
    \item \textbf{Average Holding Period (Hold):} A direct proxy for trading frequency and transaction cost efficiency. We define turnover $\tau$ as the average daily absolute change in position weights. The implied holding period is calculated relative to the average Gross Market Value (GMV) to account for leverage scaling:
    \begin{equation}
    \begin{aligned}
    \text{Hold} 
    &= \frac{2 \times 252}{\tau / \text{Avg.\ GMV}}, \\
    \text{where} \quad 
    \tau 
    &= \frac{252}{T} \sum_{t=1}^T \frac{1}{N_t} 
       \sum_{i=1}^{N_t} \lvert w_{i,t} - w_{i,t-1} \rvert .
    \end{aligned}
    \end{equation}
    This diagnostic checks whether the strategy's alpha decay matches its trading horizon. Note that all strategies are long/short, meaning positions $w_{i,t}$ can be positive (long) or negative (short).
\end{itemize}

\paragraph{Benchmark Relative Metrics.}
We compare all strategies against the \emph{Passive Equal Risk} benchmark ($R_t^{\text{bench}}$).
\begin{itemize}
    \item \textbf{Information Ratio (IR):} The risk-adjusted active return, defined as the annualized mean of the excess return spread divided by the tracking error (standard deviation of the spread):
    \begin{equation}
        \text{IR} = \sqrt{252} \cdot \frac{\widehat{\mathbb{E}}[R_t - R_t^{\text{bench}}]}{\sqrt{\widehat{\text{Var}}[R_t - R_t^{\text{bench}}]}}.
    \end{equation}
    This measures the consistency of the strategy's value-add over a passive allocation.
    
    \item \textbf{Alpha $t$-statistic ($t_\alpha$):} The HAC-adjusted $t$-statistic testing the statistical significance of the excess return ($H_0: \mathbb{E}[R_t - R_t^{\text{bench}}] \le 0$). A value $>2.0$ usually indicates statistically significant outperformance. This confirms whether the strategy provides genuine alpha beyond the risk premia available in the benchmark.
    
    \item \textbf{Correlation ($\rho$):} The Pearson correlation coefficient between the strategy's net returns and the benchmark returns:
    \begin{equation}
        \rho = \frac{\widehat{\text{Cov}}(R_t, R_t^{\text{bench}})}{\hat{\sigma}_{\text{strat}} \cdot \hat{\sigma}_{\text{bench}}}.
    \end{equation}
    Lower correlation implies the strategy captures unique structural alpha rather than static beta exposure.
\end{itemize}

\subsection{Training Protocol and Data Splits}\label{sec:experiments:splits}
We evaluate the model using a strict \emph{Walk-Forward Validation} scheme to prevent look-ahead bias. We partition the dataset (1990--2025) into five-year expanding blocks. For each block, the model is trained on all prior history, validated on the subsequent 10\% of the window, and tested on the following 5 years. The model is fully retrained every 5 years. Scalers (volatility, costs) are fit on the training set and frozen for the test period. We report performance on the union of out-of-sample blocks from 2010--2025 (inclusive).
 
To ensure robustness against initialization noise, we train 50 independent seeds for each architecture. We construct the final policy by averaging the signals of the top half ($K=25$) ranked by validation Sharpe ratio. This selection protocol (50 seeds $\to$ Top 25) was calibrated based on an out-of-sample study of the Momentum Transformer over the 2000--2010 period; the cross-sectional DeePM methods require the fuller history (training data up to 2010) to stabilize the learning of the graph structure and thus rely on the protocol established by the temporal baseline. Our results Sec.~\ref{sec:results} demonstrate that there is little difference in performance between $K\in\{10, 25,50\}$, once we have benefited from ensembling.

We use rolling sequences of length 84 trading days. Motivated by \citep{wood2024fewshot}, the first 21 steps of each sequence are treated as a burn-in period and are used solely to initialize the recurrent states, with the 63 days of evaluation aligning with the literature \citep{lim2019tsmom}. Gradients are masked during the burn-in period to ensure that optimization is driven by stable internal representations. Additional experimental details are provided in Appendix~\ref{app:exp_details}. Excluding the burn-in steps, the training and validation data are constructed by segmenting the time series into non-overlapping sequences. We use a 90/10 chronological split for training and validation. For out-of-sample backtesting, we again use sequences of length 84, but increase the burn-in period to 63 days to provide a richer historical context for trading decisions. Importantly, we do not explicitly retrain across regimes; instead, regime variation is handled implicitly through the proposed robust loss.

The SoftMin loss hyperparameters $\tau=1/5$ and $\lambda=0.1$ were selected based on validation Sharpe, which does not include the SoftMin loss term. All other hyperparameters are listed in Appendix~\ref{app:exp_details}.


\section{Empirical Findings}\label{sec:results}
\subsection{Results}
We quantify the contribution of each architectural and objective component by ablating it from the best-performing DeePM specification. The baseline is the full model trained on the proposed joint net-Sharpe objective ($\tau=0.5,\; \lambda=1/5$) with macro-graph filtering (GAT), lagged cross-sectional attention, volatility scaling, top half of seeds based on validation loss, and transaction costs with training scaler $\gamma=0.5$; it achieves a normalized out-of-sample Net Sharpe of 0.93. Each ablation modifies exactly one component. For fairness, we also train the \emph{Momentum Transformer} with transaction costs and training scaler $\gamma=0.5$.  The empirical results for the 2010--2025 period are presented in Table~\ref{tab:main_results}. 

It is important to note that our evaluation focuses on the post-2010 regime and incorporates a highly realistic transaction cost model (incorporating tick-size constraints and liquidity scalars). This rigorous setting contrasts with studies that benefit from the high-trend 1990s era or assume simplified execution costs (such as the original Momentum Transformer). Our period aligns with the ``CTA Winter'' of the 2010s, then the post-2020 volatility transition, which we further isolate in Table~\ref{tab:2020_onwards_results}, to assess resilience during the pandemic, inflation shocks, and the subsequent higher-for-longer regime.

\begin{table}[htbp]
\centering
\caption{Out-of-Sample Performance 2010--2025 (end). All strategies are rescaled to 10\% annualized volatility. \textbf{SR}: annualized Sharpe ratio. \textbf{CAGR}: compound annual growth rate. \textbf{MDD}: maximum drawdown. \textbf{Calmar}: $\mathrm{CAGR}/|\mathrm{MDD}|$. \textbf{Hold (days)}: implied average holding period in days (execution/turnover efficiency proxy; $\infty$ denotes buy-and-hold). \textbf{Gross} metrics exclude transaction costs; \textbf{Net} metrics include transaction costs. \textbf{vs Bench. (Net)}: metrics relative to Passive Equal Risk (IR: information ratio of excess return, $t_\alpha$: HAC $t$-statistic of excess return, $\rho$: return correlation). \textbf{$t$}: HAC $t$-statistic of the strategy's own net return.}

\label{tab:main_results}
\footnotesize
\setlength{\tabcolsep}{3.0pt} 
\begin{tabular}{llrrrrrrrrrr}
\toprule
& & \multicolumn{1}{c}{\textbf{Gross}} & \multicolumn{5}{c}{\textbf{Net}} & \multicolumn{1}{c}{\textbf{Exec}} & \multicolumn{3}{c}{\textbf{vs Bench. (Net)}} \\
\cmidrule(lr){3-3} \cmidrule(lr){4-8} \cmidrule(lr){9-9} \cmidrule(lr){10-12}
& \textbf{Strategy} & \textbf{SR} & \textbf{SR} & \textbf{$t$} & \textbf{CAGR} & \textbf{Calmar} & \textbf{MDD} & \textbf{(days)} & \textbf{IR} & \textbf{$t_\alpha$} & \textbf{$\rho$} \\
\midrule
\multirow{1}{*}{\rotatebox[origin=c]{90}{\textbf{}}} 
& \textbf{DeePM} & 1.29 & \textbf{0.93} & \textbf{3.69} & \textbf{9.2\%} & \textbf{0.58} & -16.0\% & 7.1 & 0.44 & 1.85 & 0.52 \\
& DeePM (MACD features) & 1.10 & 0.84 & 3.26 & 8.2\% & 0.49 & -16.9\% & 10.4 & 0.38 & 1.57 & 0.61 \\
\midrule
\midrule
\multirow{11}{*}{\rotatebox[origin=c]{90}{\textbf{Baselines}}} 
& \textbf{Passive Equal Risk (Bench.)} & 0.50 & 0.50 & 1.90 & 4.6\% & 0.17 & \underline{-27.1\%} & $\infty$ & - & - & 1.00 \\
& Trend (TSMOM) & 0.51 & 0.45 & 1.75 & 4.1\% & 0.21 & -19.8\% & 32.2 & -0.03 & -0.10 & 0.02 \\
& Risk Managed Trend & 0.49 & 0.39 & 1.50 & 3.5\% & 0.13 & -26.8\% & 14.6 & -0.07 & -0.26 & 0.02 \\
& MVO Trend & 0.55 & -0.07 & -0.26 & -1.2\% & -0.02 & -51.4\% & 5.1 & -0.41 & -1.56 & 0.05 \\
& MVO-TP Trend & 0.59 & 0.47 & 1.79 & 4.3\% & 0.15 & \underline{-28.8\%} & 15.1 & -0.01 & -0.05 & 0.06 \\
& Risk Parity Trend & 0.35 & 0.18 & 0.75 & 1.4\% & 0.04 & -32.2\% & 9.9 & -0.22 & -0.83 & 0.04 \\
& MACD Multi-Scale & 0.28 & 0.25 & 1.00 & 2.0\% & 0.08 & -23.8\% & \textbf{38.5} & -0.17 & -0.66 & \textbf{-0.05} \\
& Risk Managed MACD & 0.26 & 0.20 & 0.81 & 1.5\% & 0.05 & -27.9\% & 16.0 & -0.20 & -0.80 & -0.04 \\
& MVO-TP MACD & 0.24 & 0.15 & 0.61 & 1.0\% & 0.04 & -23.6\% & 16.6 & -0.24 & -0.97 & 0.01 \\
& Mom. Transformer ($\gamma=0$) & 1.10 & 0.60 & 2.44 & 5.6\% & 0.21 & -26.3\% & \underline{4.0} & 0.10 & 0.42 & 0.45 \\
& Mom. Transformer ($\gamma=0.5$) & 1.02 & 0.66 & 2.54 & 6.2\% & 0.20 & \underline{-31.9\%} & 5.0 & 0.15 & 0.60 & 0.39 \\
\midrule
\midrule
\multirow{7}{*}{\rotatebox[origin=c]{90}{\textbf{Architecture}}}
& Cascading lag & 1.19 & 0.84 & 3.29 & 8.2\% & 0.44 & -18.7\% & 7.5 & 0.34 & 1.41 & 0.51 \\
& Independent (No Structure)  & 1.18 & 0.83 & 3.34 & 8.2\% & 0.48 & -17.0\% & 7.7 & 0.33 & 1.35 & 0.50 \\
& No Cross-Attn (Graph Only) & 1.15 & 0.84 & 3.24 & 8.2\% & 0.44 & -18.4\% & 7.7 & 0.37 & 1.54 & 0.60 \\
& No Graph (Cross-Attn Only) & 1.13 & 0.79 & 3.11 & 7.7\% & 0.39 & -19.8\% & 7.5 & 0.29 & 1.19 & 0.51 \\
& Flip Graph/Cross-Attention & 1.21 & 0.87 & 3.38 & 8.5\% & 0.43 & -19.8\% & 7.5 & 0.39 & 1.64 & 0.58 \\
& No ReZero & 0.85 & 0.71 & 2.70 & 6.8\% & 0.40 & -17.0\% & 14.5 & 0.26 & 1.08 & 0.70 \\
& GCN (Isotropic) & 1.09 & 0.81 & 3.12 & 7.9\% & 0.44 & -17.9\% & 8.3 & 0.37 & 1.55 & 0.65 \\
\midrule
\multirow{5}{*}{\rotatebox[origin=c]{90}{\textbf{Loss}}}
& No SoftMin (Pooled Only) & 0.79 & 0.68 & 2.70 & 6.5\% & 0.50 & \textbf{-13.0\%} & 18.4 & 0.20 & 0.84 & 0.62 \\
& SoftMin $\tau=1$ & 1.25 & 0.85 & 3.40 & 8.3\% & 0.54 & -15.5\% & 6.9 & 0.34 & 1.42 & 0.49 \\
& SoftMin $\tau=0.05$ & 1.17 & 0.83 & 3.33 & 8.1\% & 0.48 & -16.7\% & 8.1 & 0.33 & 1.38 & 0.51 \\
& Zero Cost Training $\gamma=0$ & 1.17 & 0.56 & 2.29 & 5.2\% & 0.16 & \underline{-32.1\%} & 4.9 & 0.05 & 0.20 & 0.33 \\
& Full Cost Training $\gamma=1$  & 0.81 & 0.70 & 2.69 & 6.7\% & 0.41 & -16.4\% & 19.0 & 0.25 & 1.07 & \underline{0.70} \\
\midrule
\multirow{4}{*}{\rotatebox[origin=c]{90}{\textbf{Ensemble}}}
& Best Seed $K=1$ & 1.11 & 0.72 & 2.88 & 7.0\% & 0.41 & -16.8\% & 6.6 & 0.20 & 0.79 & 0.38 \\
& Top 10 Seeds $K=10$ & \textbf{1.30} & 0.93 & 3.63 & 9.1\% & 0.56 & -16.3\% & 6.7 & 0.40 & 1.67 & 0.45 \\
& All Seeds $K=50$ & 1.06 & 0.86 & 3.33 & 8.4\% & 0.57 & -14.7\% & 10.7 & 0.44 & 1.85 & 0.68 \\
& 100-seed $K=50$ (2x both) & 1.26 & 0.93 & 3.64 & 9.2\% & 0.54 & -16.9\% & 7.6 & \textbf{0.46} & \textbf{1.93} & 0.57 \\
\bottomrule
\end{tabular}
\end{table}

\begin{table}[htbp]
\centering
\caption{Out-of-Sample Performance 2020--2025 (end). All strategies are rescaled to 10\% annualized volatility. See Table~\ref{tab:main_results} for the legend.}

\label{tab:2020_onwards_results}
\footnotesize
\setlength{\tabcolsep}{3.0pt} 
\begin{tabular}{llrrrrrrrrrr}
\toprule
& & \multicolumn{1}{c}{\textbf{Gross}} & \multicolumn{5}{c}{\textbf{Net}} & \multicolumn{1}{c}{\textbf{Exec}} & \multicolumn{3}{c}{\textbf{vs Bench. (Net)}} \\
\cmidrule(lr){3-3} \cmidrule(lr){4-8} \cmidrule(lr){9-9} \cmidrule(lr){10-12}
& \textbf{Strategy} & \textbf{SR} & \textbf{SR} & \textbf{$t$} & \textbf{CAGR} & \textbf{Calmar} & \textbf{MDD} & \textbf{(days)} & \textbf{IR} & \textbf{$t_\alpha$} & \textbf{$\rho$} \\
\midrule
\multirow{1}{*}{\rotatebox[origin=c]{90}{\textbf{}}} 
& DeePM & 1.07 & 0.79 & 1.84 & 7.7\% & 0.56 & \textbf{-13.8\%} & 8.0 & 0.45 & 1.16 & 0.57 \\
& DeePM (MACD features) & \textbf{1.16} & \textbf{0.97} & \textbf{2.20} & \textbf{9.6\%} & \textbf{0.65} & -14.9\% & 11.4 & \textbf{0.68} & \textbf{1.67} & 0.62 \\
\midrule
\midrule
\multirow{11}{*}{\rotatebox[origin=c]{90}{\textbf{Baselines}}} 
& \textbf{Passive Equal Risk (Bench.)}  & 0.38 & 0.37 & 0.84 & 3.2\% & 0.17 & -18.8\% & $\infty$ & - & - & 1.00 \\
& Trend (TSMOM) & 0.43 & 0.38 & 0.89 & 3.3\% & 0.18 & -18.9\% & 24.5 & 0.00 & 0.01 & -0.08 \\
& Risk Managed Trend & 0.29 & 0.20 & 0.47 & 1.5\% & 0.07 & -21.4\% & 13.0 & -0.11 & -0.26 & -0.06 \\
& MVO Trend & 0.42 & -0.15 & -0.36 & -2.0\% & -0.06 & -34.3\% & 5.2 & -0.37 & -0.85 & 0.01 \\
& MVO-TP Trend & 0.31 & 0.20 & 0.48 & 1.6\% & 0.07 & -22.1\% & 13.1 & -0.11 & -0.26 & -0.01 \\
& Risk Parity Trend & 0.50 & 0.35 & 0.84 & 3.1\% & 0.12 & -25.7\% & 8.9 & -0.01 & -0.02 & -0.08 \\
& MACD (Multi-Scale) & 0.29 & 0.26 & 0.65 & 2.2\% & 0.09 & -24.4\% & \textbf{35.9} & -0.07 & -0.17 & \textbf{-0.09} \\
& Risk Managed MACD & 0.10 & 0.04 & 0.09 & -0.1\% & -0.00 & -27.2\% & 15.7 & -0.23 & -0.52 & -0.08 \\
& MVO-TP MACD & 0.11 & 0.03 & 0.08 & -0.2\% & -0.01 & -23.3\% & 16.3 & -0.23 & -0.56 & -0.04 \\
& Mom. Transformer ($\gamma=0$) & 0.64 & 0.18 & 0.45 & 1.3\% & 0.07 & -18.6\% & 3.8 & -0.19 & -0.49 & 0.52 \\
& Mom. Transformer ($\gamma=0.5$) & 0.65 & 0.38 & 0.87 & 3.3\% & 0.16 & -20.7\% & 5.7 & 0.01 & 0.02 & 0.62 \\
\bottomrule
\end{tabular}
\end{table}

\begin{figure}[htbp]
  \centering
  \includegraphics[width=\textwidth]{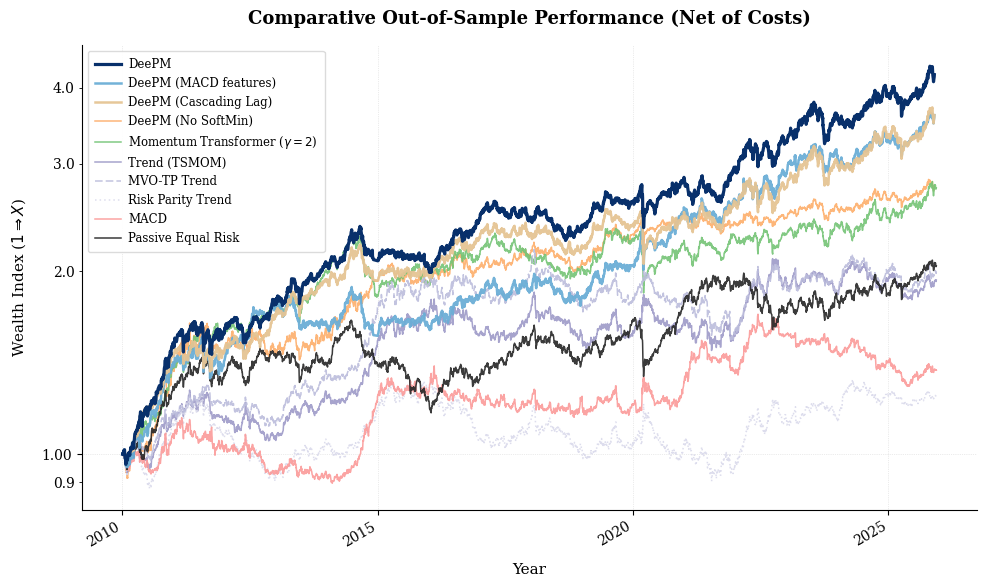} 
  \caption{Cumulative net-of-cost wealth growth for DeePM variants versus standard systematic baselines (2010--2025). The y-axis utilizes a logarithmic scale to properly visualize long-term compounding differences.}
  \label{fig:pnl_comparison}
\end{figure}

\subsection{Discussion of Empirical Findings}\label{sec:results:discussion}

\paragraph{\textbf{Economic Significance and Net Performance.}}
The DeePM Ensemble demonstrates statistically significant outperformance against both passive risk premia and deep learning baselines. Notably, the proposed model achieves a Net Sharpe ratio of \textbf{0.93} with a highly significant HAC $t$-statistic of \textbf{3.69}, confirming the existence of a robust risk premium net of transaction costs. This performance exceeds the Passive Equal Risk (0.50) and TSMOM (0.45) benchmarks by a wide margin. Crucially, it outperforms the state-of-the-art Momentum Transformer trained with the same transaction cost regularization ($\gamma=0.5$) which achieved a Net Sharpe of 0.66, highlighting the specific contribution of the structural graph prior. The strategy achieves an Information Ratio of 0.44 relative to the passive benchmark with an average holding period of 7.1 days, indicating that the model successfully identifies transient structural alpha opportunities distinct from static factor exposure ($t_\alpha = 1.85$). This margin is particularly notable given that the 2010--2025 test set corresponds to a historically exceptional regime of sustained asset appreciation (often termed an ``Everything Bubble''), making passive risk-parity a uniquely challenging baseline to beat net of costs. Furthermore, the extended 100-seed ensemble ($K=50$) achieves a Net Information Ratio $t$-statistic of 1.93 (approximate $p$-value 0.05), confirming statistical significance over this outlier passive history. The convergence between Gross (1.29) and Net (0.93) Sharpe ratios suggests that the direct optimization of transaction-adjusted returns effectively internalizes execution constraints.

\paragraph{\textbf{Role of the Two Spatial Inductive Biases.}}
The ablation study confirms a critical interaction between the topological (Graph) and data-driven (Cross-Attention) components. Notably, neither mechanism succeeds in isolation. The \emph{No Graph} variant (Cross-Attn Only) performs worse than the independent baseline (Net Sharpe 0.79 vs 0.83), suggesting that unconstrained attention overfits spurious correlations. Conversely, the \emph{Graph Only} variant (Net Sharpe 0.84) offers marginal improvement, implying that static economic priors are insufficient without dynamic weighting. The full DeePM model (Net Sharpe 0.93) outperforms both. Furthermore, the ordering of these modules is significant; reversing the sequence (Graph then Cross-Attention) degrades performance to a Net Sharpe of 0.87. This supports the hypothesis that the economic graph functions best as a regularizing filter applied \textit{after} the model has learned raw cross-sectional interactions, effectively ``denoising'' the data-driven signals and reducing Maximum Drawdown from 19.8\% (flipped) to 16.0\%.

\paragraph{\textbf{Portfolio-Centric Supervision.}}
The results further highlight the superiority of portfolio-level supervision over asset-level objectives. Even the ``Independent (No Structure)'' ablation, which treats assets in isolation, achieves a Net Sharpe of 0.83, comfortably outperforming the ``Momentum Transformer'' baseline (0.66). Since both models utilize similar temporal encoders, this confirms that, with suitable optimization stability, aligning the loss function with the ultimate trading objective provides a superior optimization landscape, even in the absence of explicit cross-sectional modelling.

\paragraph{\textbf{Information Latency and Causal Validity.}}
Comparing the ``Directed Delay'' protocol against the ``Cascading Lag'' variant reveals a preference for causal robustness over information freshness. The model utilizing strictly lagged data ($t-1$) outperforms the variant using same-day data from earlier-closing markets (Sharpe 0.93 vs. 0.84). This finding implies that maximizing information freshness introduces non-stationary intraday noise that degrades generalization. By enforcing a strict delay, the model is compelled to learn persistent impulse response functions and transfer entropy rather than transient co-movements.

\paragraph{\textbf{Optimization Stability and Addressing the Inertia Trap.}}
Our ablation study identifies the choice of objective function as the single most impactful component of the proposed framework, exceeding even the contribution of the graph-based architectural priors. While removing the macroeconomic graph structure degrades performance from 0.93 to 0.79 (-15\%), reverting from the Robust SoftMin to a standard Pooled Sharpe objective causes a catastrophic drop to 0.68 (-27\%). This hierarchy suggests that while spatial structure is beneficial, the primary challenge in financial deep learning is not merely representation learning, but optimization stability. We observed that the SoftMin loss significantly stabilized training dynamics, reducing the divergence between training and validation loss curves.

This stability is critical for complex, ratio-based objectives like the Sharpe ratio, which are prone to collapsing into degenerate local minima where the model minimizes the denominator (volatility) rather than maximizing the numerator (return). A counter-intuitive artifact of this phenomenon is that the \emph{No SoftMin} ablation exhibits the lowest Maximum Drawdown (-13.0\%). This is not a sign of robustness, but of ``inertia'': without the worst-window penalty (approximating Entropic Value-at-Risk), the model retreats to a low-frequency holding pattern (18.4 days) to minimize costs, effectively becoming a slow-moving beta strategy. In contrast, the robust SoftMin objective compels the agent to trade actively (7.1 days) to defend performance during adverse regimes, accepting marginally higher volatility for vastly superior risk-adjusted returns (Sharpe 0.93).

\paragraph{\textbf{Adversarial Temperature.}}
We further analyse the role of the adversarial temperature $\tau$, which controls the ``paranoia'' of the SoftMin operator (approximating EVaR). The baseline DeePM (optimized $\tau$) outperforms both the looser $\tau=1$ specification (Net Sharpe 0.85) and the strictly harder minimax limit $\tau=0.05$ (Net Sharpe 0.83). This confirms that a calibrated degree of robustness is required: $\tau=0.05$ approximates a hard-minimax objective that can lead to optimization instability by overfitting to the single worst historical window, while  $\tau=1$ is closer to a simple average that fails to sufficiently penalize tail risks. The intermediate $\tau=0.2$ enables the model to focus on the tail of the distribution without discarding the signal from the body.

\begin{figure}[htbp]
  \centering
  \includegraphics[width=\textwidth]{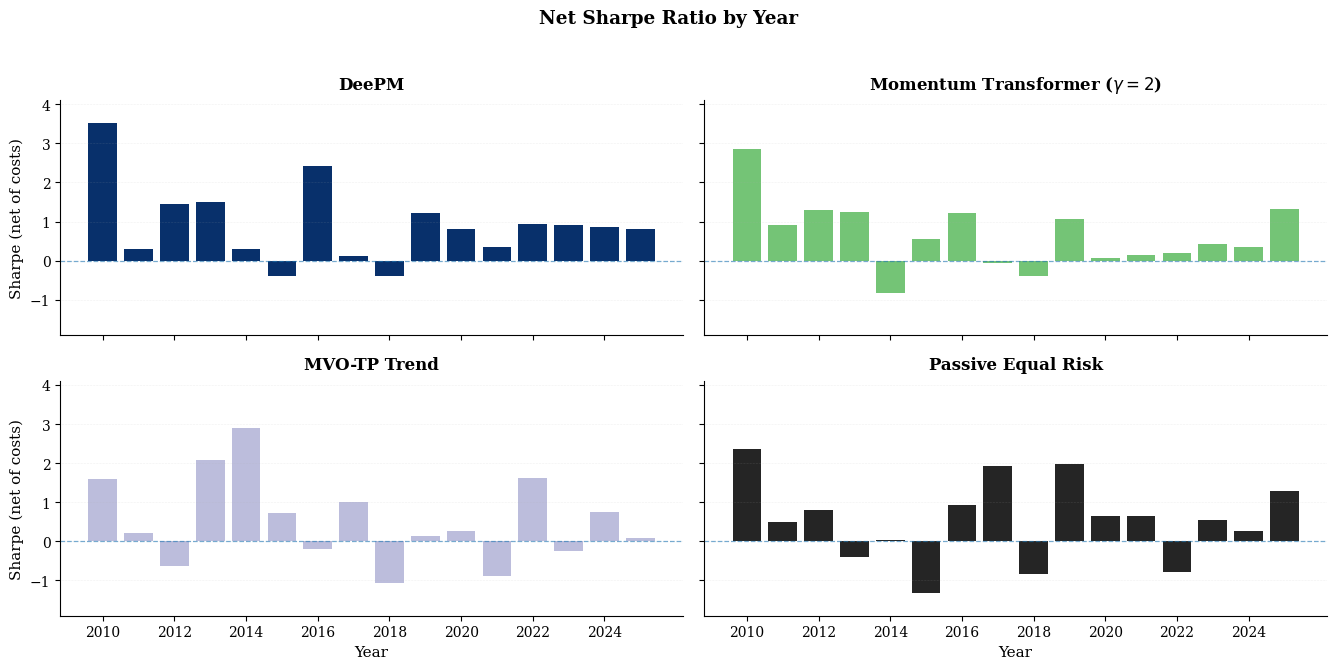} 
  \caption{Rolling 12-month Sharpe Ratio of DeePM versus the TSMOM baseline. The proposed model exhibits superior stability, maintaining positive performance during periods where classical trend-following suffers significant drawdowns (e.g., 2016, post-2020).}
  \label{fig:regime_robustness}
\end{figure}

\paragraph{\textbf{Regime Robustness and Stability.}}
Figure~\ref{fig:regime_robustness} illustrates the rolling performance stability of the DeePM Ensemble relative to classical baselines. A key deficiency of traditional trend-following (TSMOM) is its binary performance profile -- it excels in sustained divergence but suffers excessively during mean-reverting or sideways markets. In contrast, DeePM demonstrates remarkable consistency across disparate regimes. By effectively navigating diverse market environments, the model proves that its learned non-linear signal processing can identify profitable opportunities even when simple persistent trends are absent, reducing the ``feast-or-famine'' cycle typical of systematic macro strategies.

This resilience is most distinct during the post-2020 block (Table~\ref{tab:2020_onwards_results}), where the model delivered sustained alpha despite the transition from pandemic-induced volatility to inflationary rate shocks. While classical Trend strategies faltered (Sharpe 0.38) and passive allocation struggled (Sharpe 0.37) during this period, the DeePM framework maintained a Net Sharpe of 0.79 (0.97 with MACD features), suggesting the learned representations successfully generalized from the low-rate training era to the new high-rate regime without requiring retraining.

\paragraph{\textbf{Cost-Aware Learning.}}
The results validate the necessity of end-to-end cost optimization. The convergence between Gross (1.29) and Net (0.93) Sharpe ratios suggests that the direct optimization of transaction-adjusted returns effectively internalizes execution constraints. Training with \emph{Zero Cost} ($\gamma=0$) leads to over-trading (Hold 4.9 days) and a collapse in Net Sharpe to 0.56; without a penalty, the model aggressively fits high-frequency noise that appears profitable in frictionless theory but is illusory in practice. This sensitivity is markedly more acute here than in univariate frameworks like the Momentum Transformer. In a portfolio context, a zero-cost objective encourages the model to exploit spurious cross-sectional arbitrages (e.g., long Asset A vs. short Asset B based on transient noise), creating a much larger surface area for unmonetizable turnover than simple directional trend following. Conversely, \emph{Full Cost} training ($\gamma=1$) results in an overly conservative policy (Hold 19.0 days, Sharpe 0.70). The optimal performance (Sharpe 0.93) is achieved with an intermediate penalty, confirming theoretical predictions that ensembling provides implicit regularization, allowing the explicit penalty to be relaxed to capture higher-frequency alpha.

\paragraph{\textbf{Limitations of Two-Stage Mean-Variance Frameworks.}}
The comparison with MVO baselines highlights the advantages of an end-to-end approach over separated estimation and optimization. The standard two-stage MVO fails catastrophically (Sharpe -0.07) due to error maximization, where estimation noise in the covariance matrix is amplified by the optimizer. While introducing a quadratic turnover penalty (MVO-TP) stabilizes the strategy and restores positive performance (Sharpe 0.47), it remains significantly inferior to the deep learning approach. This suggests that the performance gap is not merely a function of turnover management, but stems from DeePM's ability to learn non-linear predictive signals that two-stage linear models cannot capture.

\paragraph{\textbf{Ensembling as Structural Regularization.}}
The results confirm that ensembling acts as an effective variance reduction technique for portfolio construction. The ensemble model improves the Net Sharpe ratio from 0.72 (single best seed) to 0.93 (Top 10 seeds). Crucially, performance remains remarkably stable beyond this initial uplift: the Baseline ($K=25$), Top 10 ($K=10$), and the extended 100-seed ensemble ($K=50$) all converge to a Net Sharpe of approximately 0.93. This insensitivity indicates that the specific ensemble size is not a cherry-picked hyperparameter, provided a sufficient diversity of models is aggregated. By averaging over multiple initializations, the ensemble mitigates the idiosyncrasies of individual training runs, resulting in a smoother position path that is more efficient to execute and stable during drawdowns.

\section{Conclusions and Future Work}\label{sec:conclusion}
This paper introduces DeePM, an end-to-end portfolio management framework that integrates macroeconomic structural priors with deep representation learning. Our empirical results demonstrate that imposing a static sector-graph structure acts as a vital regularizer, enabling the model to learn robust cross-sectional alpha that generalizes out-of-sample (Net Sharpe 0.93) where purely data-driven attention mechanisms fail (Net Sharpe 0.79). Furthermore, we show that the choice of loss function is paramount; the robust SoftMin (EVaR) objective significantly outperforms standard mean-variance and pooled objectives by preventing overfitting to low-volatility regimes and avoiding the ``inertia trap.'' Beyond architectural novelty, this work bridges the gap between academic deep learning and institutional practice. By rigorously modelling transaction costs, enforcing filtration-compliant delays for asynchronous global markets, and optimizing for regime-robust metrics rather than average-case performance, DeePM offers a deployable blueprint for systematic macro managers seeking to modernize legacy trend-following systems.

Future research will focus on five directions: (1) \textit{Dynamic Graph Learning}, allowing the adjacency matrix to evolve over time rather than relying on fixed sector definitions; (2) \textit{Interpretability}, leveraging GNN explainability techniques to map learned attention weights back to specific economic transmission channels; (3) \textit{Hierarchical Structure}, extending the graph prior to explicitly model multi-level relationships across assets, sectors, and asset classes; (4) \textit{Expanded Information Sets}, integrating carry and macroeconomic data to capture drivers orthogonal to pure price momentum; (5) \textit{Freshness vs. Causality}, benchmarking against a fully synchronous data source to rigorously test whether the ``Causal Sieve'' benefit of the Directed Delay protocol outweighs the informational cost of sacrificing ``instantaneous'' data freshness; and (6) \textit{High-Frequency Generalization}, extending the framework to intraday data to test whether the proposed inductive biases persist at higher frequencies, and determining the time-scale at which the trade-off between information freshness and causal validity inverts.

\section{Acknowledgements}
KW would like to thank the Oxford-Man Institute of Quantitative Finance for its generous support.
SR would like to thank the U.K. Royal Academy of Engineering.

\bibliographystyle{plainnat}
\bibliography{references}

\appendix
\section{Universe Specifications and Methodology}\label{app:universe}

This appendix details the investment universe, the macroeconomic graph topology used for structural regularization, and the transaction cost models applied during backtesting.

\subsection{Data Provenance and Processing}\label{app:universe:data}
We source historical daily open, high, low, and close (OHLC) prices for all 50 futures and FX contracts from the Pinnacle Data Corp CLC Database \citep{pinnacledata}; however, for all experiments we just use close to construct features. To construct continuous return series from individual contract expirations, we utilize ratio-adjusted (``Panama'') continuous contracts.

Unlike additive adjustment methods, which shift historical prices by a fixed absolute amount and can lead to negative prices or distorted percentage returns over long horizons, ratio adjustment scales historical prices by the ratio of the new contract price to the old contract price at each roll date. This methodology rigorously preserves the relative percentage returns and volatility structure of the asset series, ensuring that the inputs to the volatility-scaling mechanism (Section 3.2) and the training returns for the network faithfully represent the realized distribution of market returns.

\subsection{Macro Graph Topology}\label{app:universe:graph}
The macroeconomic prior graph $\mathcal{G} = (\mathcal{V}, \mathcal{E})$ is constructed deterministically based on economic first principles. The adjacency matrix $A$ is formed via the union of the following edge sets:

\begin{enumerate}
    \item \textbf{Intra-Group Cliques (Sectoral Homophily):} All nodes within a specific Macro Group (e.g., \texttt{COMM\_ENERGY}) are fully connected, enforcing the view that assets within a sub-sector share a single latent factor.
    \item \textbf{Risk-On Channel:} Connects Global Equities $\leftrightarrow$ Base Metals $\leftrightarrow$ Risk FX (AUD, CAD, MXN).
    \item \textbf{Inflation Channel:} Connects Energy $\leftrightarrow$ US Treasuries $\leftrightarrow$ Precious Metals.
    \item \textbf{Safe Haven Flows:} Connects US Treasuries $\leftrightarrow$ Safe FX (JPY, CHF) $\leftrightarrow$ Precious Metals.
    \item \textbf{Commodity Exporters:} Connects Energy/Metals to their respective currency proxies (e.g., Oil $\leftrightarrow$ CAD/MXN, Metals $\leftrightarrow$ AUD).
    \item \textbf{Regional Integration:} Creates a triangular linkage between the primary Equity index, Sovereign Bond, and Currency for major economic zones (US, EU, JP, UK, CA).
\end{enumerate}

\subsection{Transaction Cost Methodology}\label{app:universe:tcost}
We implement a \emph{Structural Minimum Cost Model} that synthesizes the ``Tick Size Constraint" theory (Harris, 2003) with market impact estimates. The cost $c_i$ (in basis points) is derived as:
\begin{equation}
    c_i \approx \max(C_{\text{floor}}, \; C_{\text{struct}, i} \times \lambda_i)
\end{equation}
where $C_{\text{struct}}$ is the theoretical cost implied by the minimum price variation (tick size) relative to the contract value, and $\lambda_i \ge 1.0$ is a liquidity scalar. 

\subsubsection{Justification of Liquidity Scalars ($\lambda_i$)}
While liquid electronic markets (e.g., S\&P 500, US Treasuries) typically trade near their structural tick constraints ($\lambda_i \approx 1.0-1.5$), we apply significant scalers to subsets of the universe to reflect institutional execution realities that simple tick-based models miss.

\textbf{The ``Roach Motel" Effect (Depth vs. Spread).}
Certain markets, such as Palladium (\texttt{PA}) or specific Softs, may exhibit tight top-of-book spreads that mask a lack of order book depth. For institutional-size execution, the ``effective spread" to sweep the book is significantly wider than the quoted spread. We model this via high ``Structure" scalars (e.g., Palladium $\lambda_i \approx 2.4$, Orange Juice $\lambda_i \approx 12.6$) to penalize strategies that assume they can exit large positions at the touch price during stress periods.

\textbf{Volatility-Constrained Market Making.}
For assets like Natural Gas or Livestock (Feeder Cattle), the exchange-mandated tick size is often non-binding. Market makers widen spreads to compensate for extreme inventory volatility. In these regimes, the cost is driven by volatility rather than quantization noise. We apply scalars ($\lambda_i > 2.0$) to align the modeled cost with the realized volatility-adjusted spreads observed in 2023-2024.

\textbf{Regional Session Impacts.}
Assets such as the Nikkei 225 often show adequate liquidity during their local session but suffer from ``air pockets" during the US/EU overlap when global macro rebalancing occurs. We apply an ``Impact" scalar ($\lambda_i \approx 2.0-2.3$) to represent the higher slippage associated with trading these assets out of their primary liquidity window.

\textbf{High-Velocity and Arbitrage Discounts ($\lambda_i < 1.0$).}
Conversely, certain highly efficient markets (e.g., Hang Seng, Dollar Index, FTSE 100) are assigned scalars below unity. In these cases, the exchange-mandated tick size ($C_{\text{struct}}$) acts as a legacy constraint that overestimates the effective cost for institutional participants. Deep spot-futures arbitrage links (e.g., between the Dollar Index future and the underlying spot FX basket) allow market makers to quote aggressive effective spreads, often facilitating execution better than the naked tick implies via block liquidity or EFP (Exchange for Physical) mechanisms.

\subsubsection{Cost Summary Statistics}
Table \ref{tab:summary_stats} summarizes the resulting cost distribution. The divergence between ``Median Calc" and the ``Final Band" in the high-cost tiers reflects the necessity of the $\lambda_i$ scalars.

\begin{table}[htbp]
    \centering
    \footnotesize
    \caption{Transaction Cost Regimes: Theoretical vs. Modeled}
    \label{tab:summary_stats}
    \begin{tabular}{lcccl}
        \toprule
        \textbf{Final Band} & \textbf{Median $C_{struct}$} & \textbf{Mean $C_{struct}$} & \textbf{Mean $\lambda_i$} & \textbf{Typical Asset Class} \\
        \midrule
        0.25 bps & 0.19 & 0.19 & 1.6x & Ultra-Liquid (S\&P 500, US Treasuries) \\
        0.50 bps & 0.40 & 0.43 & 1.5x & Very Liquid (Bund, JPY, FTSE) \\
        0.75 bps & 0.71 & 0.81 & 1.1x & Liquid Physicals (Crude, 10Y Note) \\
        1.00 bps & 1.02 & 0.96 & 1.1x & Standard (Silver, EuroStoxx) \\
        1.50 bps & 1.30 & 1.02 & 1.6x & Mid-Liquidity (30Y Bond, Platinum) \\
        2.50 bps & 1.67 & 1.60 & 1.9x & High Cost (Grains, Nat Gas, Livestock) \\
        6.00 bps & 1.75* & 1.75* & 4.2x & Volatile / Thin (Palladium, Cocoa) \\
        15.0 bps & 1.19* & 1.19* & 12.6x & Distressed (Orange Juice) \\
        \bottomrule
    \end{tabular}
    \par\medskip
    \textit{*Note: For high-cost bands, the divergence between ``Median Calc" and the Final Band reflects the high Liquidity Scalar required for volatile markets.}
\end{table}

\subsection{Master Data Universe} \label{app:universe:master}
Table \ref{tab:master_universe} provides the granular derivation of costs for the full 50-asset universe.
\begin{itemize}
    \item \textbf{Calc:} The theoretical cost based on Tick Size ($C_{struct}$).
    \item \textbf{Scalar ($\lambda_i$):} The multiplier applied ($Final = Calc \times \lambda_i$).
    \item \textbf{Type:} \textbf{Impact} (Size/Session adjustment) or \textbf{Structure} (Microstructure/Volatility override).
\end{itemize}

{\scriptsize
\setlength{\tabcolsep}{3pt}
\begin{longtable}{lllcccl}
    \caption{Master Universe: Macro Groups and Transaction Cost Derivation} \label{tab:master_universe} \\
    \toprule
    \textbf{Ticker} & \textbf{Name} & \textbf{Group} & \textbf{Calc} & \textbf{$\bm{\lambda_i}$} & \textbf{Final} & \textbf{Type / Note} \\
    \midrule
    \endfirsthead
    \toprule
    \textbf{Ticker} & \textbf{Name} & \textbf{Group} & \textbf{Calc} & \textbf{$\bm{\lambda_i}$} & \textbf{Final} & \textbf{Type / Note} \\
    \midrule
    \endhead
    \bottomrule
    \endfoot

    \multicolumn{7}{l}{\textit{\textbf{\underline{Sovereign Rates}}}} \\
    TU & US 2yr Note & RATES\_US & 0.19 & 1.3x & \textbf{0.25} & Ultra-Liquid \\
    FV & US 5yr Note & RATES\_US & 0.25 & 1.0x & \textbf{0.25} & Ultra-Liquid \\
    TY & US 10yr Note & RATES\_US & 0.71 & 1.1x & \textbf{0.75} & Benchmark \\
    US & US 30yr Bond & RATES\_US & 1.32 & 1.1x & \textbf{1.50} & Liquid \\
    DU & Euro Schatz & RATES\_EU & 0.25 & 1.0x & \textbf{0.25} & Short-End EU \\
    OE & German Bobl & RATES\_EU & 1.32 & 1.1x & \textbf{1.50} & Mid-Curve EU \\
    RX & Euro Bund & RATES\_EU & 0.25 & 2.0x & \textbf{0.50} & \textbf{Impact} (Duration) \\
    G & Long Gilt & RATES\_OTHR & 0.25 & 2.0x & \textbf{0.50} & \textbf{Impact} (Non-US) \\
    CN & Canada 10yr & RATES\_OTHR & 0.25 & 2.0x & \textbf{0.50} & \textbf{Impact} (Depth) \\

    \multicolumn{7}{l}{\textit{\textbf{\underline{Equities}}}} \\
    ES & S\&P 500 & EQUITY\_US & 0.18 & 1.4x & \textbf{0.25} & Global Benchmark \\
    EN & Nasdaq 100 & EQUITY\_US & 0.05 & 5.0x & \textbf{0.25} & Floor Effect \\
    YM & Dow Jones & EQUITY\_US & 0.10 & 2.5x & \textbf{0.25} & Floor Effect \\
    RTY & Russell 2000 & EQUITY\_US & 0.18 & 2.8x & \textbf{0.50} & Liquid Small Cap \\
    VG & EuroStoxx 50 & EQUITY\_EU & 1.04 & 1.0x & \textbf{1.00} & Standard EU Liq \\
    Z & FTSE 100 & EQUITY\_EU & 0.65 & 0.8x & \textbf{0.50} & Developed Mkt \\
    CF & CAC 40 & EQUITY\_EU & 0.65 & 0.8x & \textbf{0.50} & Developed Mkt \\
    NK & Nikkei 225 & EQUITY\_APAC & 0.64 & 2.3x & \textbf{1.50} & \textbf{Impact} (Asian Session) \\
    HI & Hang Seng & EQUITY\_APAC & 1.35 & 0.6x & \textbf{0.75} & High Velocity \\

    \multicolumn{7}{l}{\textit{\textbf{\underline{Foreign Exchange}}}} \\
    DX & Dollar Index & FX\_G10 & 0.65 & 0.8x & \textbf{0.50} & Spot Arb \\
    EU & EUR/USD & FX\_G10 & 0.24 & 1.0x & \textbf{0.25} & Global Anchor \\
    JY & JPY/USD & FX\_G10 & 0.39 & 1.3x & \textbf{0.50} & \textbf{Impact} Scaling \\
    BP & GBP/USD & FX\_G10 & 0.39 & 1.3x & \textbf{0.50} & Standard G10 \\
    CD & CAD/USD & FX\_G10 & 0.36 & 1.4x & \textbf{0.50} & Standard G10 \\
    AD & AUD/USD & FX\_G10 & 0.40 & 1.3x & \textbf{0.50} & Standard G10 \\
    SF & CHF/USD & FX\_G10 & 0.50 & 1.0x & \textbf{0.50} & Standard G10 \\
    PE & Mexican Peso & FX\_EM & 1.02 & 1.0x & \textbf{1.00} & High Vol/Low Price \\

    \multicolumn{7}{l}{\textit{\textbf{\underline{Commodities: Energy}}}} \\
    CL & WTI Crude & COMM\_EN & 0.71 & 1.1x & \textbf{0.75} & Global Benchmark \\
    CO & Brent Crude & COMM\_EN & 0.71 & 1.1x & \textbf{0.75} & Global Benchmark \\
    XB & RBOB Gas & COMM\_EN & 1.30 & 1.2x & \textbf{1.50} & Product Spread \\
    QS & Gasoil & COMM\_EN & 1.30 & 1.2x & \textbf{1.50} & Product Spread \\
    NG & Natural Gas & COMM\_EN & 1.67 & 1.5x & \textbf{2.50} & \textbf{Structure} (Vol) \\

    \multicolumn{7}{l}{\textit{\textbf{\underline{Commodities: Metals}}}} \\
    GC & Gold & COMM\_PREC & 0.19 & 1.3x & \textbf{0.25} & Tight Book \\
    SI & Silver & COMM\_PREC & 0.81 & 1.2x & \textbf{1.00} & Standard \\
    PL & Platinum & COMM\_PREC & 0.53 & 2.8x & \textbf{1.50} & \textbf{Structure} (Wide) \\
    PA & Palladium & COMM\_PREC & 2.50 & 2.4x & \textbf{6.00} & \textbf{Structure} (Thin) \\
    HG & Copper & COMM\_BASE & 0.58 & 1.3x & \textbf{0.75} & High Efficiency \\

    \multicolumn{7}{l}{\textit{\textbf{\underline{Commodities: Agriculture \& Livestock}}}} \\
    C & Corn & COMM\_AGRI & 2.91 & 0.9x & \textbf{2.50} & Standard Grain \\
    S & Soybeans & COMM\_AGRI & 1.26 & 1.2x & \textbf{1.50} & Deep Market \\
    SM & Soybean Meal & COMM\_AGRI & 1.72 & 1.5x & \textbf{2.50} & \textbf{Impact} (Crush) \\
    BO & Soybean Oil & COMM\_AGRI & 1.30 & 1.2x & \textbf{1.50} & Crush Spread \\
    W & Wheat & COMM\_AGRI & 2.27 & 1.1x & \textbf{2.50} & Volatile \\
    KW & KC Wheat & COMM\_AGRI & 2.50 & 1.0x & \textbf{2.50} & Volatile \\
    SB & Sugar & COMM\_SOFT & 2.38 & 1.1x & \textbf{2.50} & Standard Soft \\
    KC & Coffee & COMM\_SOFT & 0.81 & 3.1x & \textbf{2.50} & \textbf{Structure} (Vol) \\
    CC & Cocoa & COMM\_SOFT & 1.00 & 6.0x & \textbf{6.00} & \textbf{Structure} (Extreme) \\
    CT & Cotton & COMM\_SOFT & 0.70 & 3.6x & \textbf{2.50} & \textbf{Structure} (Thin) \\
    JO & Orange Juice & COMM\_SOFT & 1.19 & 12.6x & \textbf{15.0} & \textbf{Structure} (Distressed) \\
    LC & Live Cattle & COMM\_LIVE & 0.68 & 2.2x & \textbf{1.50} & \textbf{Structure} (Gappy) \\
    FC & Feeder Cattle & COMM\_LIVE & 0.49 & 5.1x & \textbf{2.50} & \textbf{Structure} (Gappy) \\
    LH & Lean Hogs & COMM\_LIVE & 1.47 & 1.7x & \textbf{2.50} & \textbf{Structure} (Gappy) \\
\end{longtable}
}

\subsection{Specific Overrides}
We apply manual overrides where the theoretical tick cost underestimates institutional execution difficulty:
\begin{itemize}
    \item \textbf{Floor Effect:} Liquid indices like Nasdaq 100 have a structural cost $<0.1$ bps. We floor this at 0.25 bps to represent clearing/execution minimums.
    \item \textbf{``Roach Motel" Liquidity:} Palladium is theoretically cheap but lacks depth. We scale it to 6.0 bps.
    \item \textbf{Distressed Liquidity:} Orange Juice is assigned 15.0 bps to reflect ``widowmaker" liquidity risks.
\end{itemize}

\section{Turnover Guarantees for Ensembles}\label{app:turnover_proofs}

This appendix provides formal derivations regarding the convexity of the transaction cost objective and the regularization properties of the ensemble.

\subsection{Convexity of Turnover and Ensemble Guarantee}

\begin{proposition}[Convexity of Turnover Cost]
Let $\bm{p}_{0:L}$ be a sequence of portfolio position vectors with $\bm{p}_t\in\mathbb{R}^N$. 
Assume the scaling vectors satisfy $\bm{v}_t \in \mathbb{R}^N_{+}$ and are exogenous (i.e., do not depend on $\bm{p}$). 
Define
\[
\mathcal{C}(\bm{p}) \;=\; \gamma \sum_{t=1}^L \big\| \bm{v}_t \odot (\bm{p}_t - \bm{p}_{t-1}) \big\|_1,
\qquad \gamma \ge 0.
\]
Then $\mathcal{C}(\bm{p})$ is convex in $\bm{p}_{0:L}$.
\end{proposition}

\begin{proof}
For each $t$, the map $\bm{p}_{0:L} \mapsto \bm{p}_t-\bm{p}_{t-1}$ is linear. 
Elementwise scaling by a fixed $\bm{v}_t$ is also linear, since $\bm{v}_t \odot \bm{x} = \mathrm{Diag}(\bm{v}_t)\bm{x}$. 
Thus the composition $\bm{p}\mapsto \bm{v}_t \odot (\bm{p}_t-\bm{p}_{t-1})$ is linear. 
Because $\|\cdot\|_1$ is convex and convexity is preserved under composition with an affine map, 
$\bm{p}\mapsto \| \bm{v}_t \odot (\bm{p}_t-\bm{p}_{t-1})\|_1$ is convex. 
Summing over $t$ with nonnegative weight $\gamma\ge 0$ preserves convexity, hence $\mathcal{C}$ is convex.
\end{proof}

\begin{corollary}[Ensembling Reduces Cost (Executed Mean Policy)]
Let $\{\bm{p}^{(k)}\}_{k=1}^K$ be $K$ policies and define the executed mean policy 
$\bar{\bm{p}} := \frac{1}{K}\sum_{k=1}^K \bm{p}^{(k)}$ (averaged at the \emph{position} level).
Then, by convexity of $\mathcal{C}$,
\begin{equation}
    \mathcal{C}(\bar{\bm{p}}) \le \frac{1}{K}\sum_{k=1}^K \mathcal{C}(\bm{p}^{(k)}).
\end{equation}
\end{corollary}

\begin{proof}
Since $\mathcal{C}$ is convex, Jensen's inequality gives
\[
\mathcal{C}\!\left(\frac{1}{K}\sum_{k=1}^K \bm{p}^{(k)}\right) \le \frac{1}{K}\sum_{k=1}^K \mathcal{C}\!\left(\bm{p}^{(k)}\right),
\]
which is the desired result.
\end{proof}

\noindent\textit{Note:} This guarantee applies when the ensemble is implemented by executing the averaged positions $\bar{\bm{p}}$.
If instead each policy $\bm{p}^{(k)}$ is traded separately and P\&L is averaged ex post, transaction costs are incurred per-policy and the inequality need not reflect realized costs.

\subsection{Optimal Regularization under Ensembling}
\label{prop:half-penalty}

\begin{proposition}[Optimal Penalty Decreases with Ensemble Size]
Let $\gamma \ge 0$ be the explicit turnover penalty strength used during training, and let $K$ be the ensemble size. 
We model the realized Net Sharpe Ratio as
\[
\mathrm{SR}_{\mathrm{net}}(\gamma, K) \;=\; \mathrm{SR}_{\mathrm{gross}}(\gamma) \;-\; \rho(K)\,\Delta(\gamma),
\]
where $\Delta(\gamma)$ is the turnover cost and $\rho(K)$ is an ensemble-dependent cost reduction factor.

Under the standard assumptions that:
\begin{enumerate}
    \item \textbf{Gross performance is strictly concave in regularization} ($\mathrm{SR}''_{\mathrm{gross}}(\gamma) < 0$): 
    Intuitively, relaxing constraints yields diminishing returns; the initial units of freedom capture high-conviction opportunities, while further relaxation allows the model to chase increasingly marginal and noisy signals. 
    
    \item \textbf{Turnover cost is strictly decreasing and convex in regularization} ($\Delta'(\gamma) < 0, \Delta''(\gamma) > 0$): 
    This reflects the fact that a small penalty is sufficient to eliminate the most expensive high-frequency ``noise trading,'' while further increases yield progressively smaller reductions in turnover as the portfolio approaches a static hold. 
    
    \item \textbf{Ensembling reduces realized turnover costs} ($\rho'(K) < 0$): 
    As discussed in Sec.~\ref{sec:objective:ensembling}, independent models often make uncorrelated errors, averaging their positions cancels out idiosyncratic ``noise trades,'' naturally dampening turnover without requiring explicit penalties. 
\end{enumerate}
and assuming an interior maximizer exists, the optimal training penalty $\gamma^\star(K)$ is strictly decreasing in the ensemble size $K$.
\end{proposition}

\begin{lemma}[Sufficient condition for an interior optimizer on $(0,1)$]
Assume $\gamma \in [0,1]$ and $\Phi(\gamma,\rho)$ is continuously differentiable in $\gamma$.
If
\[
\Phi_\gamma(0,\rho) > 0
\quad\text{and}\quad
\Phi_\gamma(1,\rho) < 0,
\]
then there exists $\gamma^\star \in (0,1)$ such that $\Phi_\gamma(\gamma^\star,\rho)=0$.
\end{lemma}

\begin{proof}
By continuity of $\Phi_\gamma(\cdot,\rho)$ and the Intermediate Value Theorem, it must cross zero on $(0,1)$.
\end{proof}

\noindent\textit{Specialization to our objective.}
Here $\Phi(\gamma,\rho) = \mathrm{SR}_{\mathrm{gross}}(\gamma) - \rho \Delta(\gamma)$ and hence
\[
\Phi_\gamma(\gamma,\rho)=\mathrm{SR}'_{\mathrm{gross}}(\gamma)-\rho\Delta'(\gamma).
\]
Therefore a sufficient condition for $\gamma^\star\in(0,1)$ is
\[
\mathrm{SR}'_{\mathrm{gross}}(0)-\rho\Delta'(0) > 0
\quad\text{and}\quad
\mathrm{SR}'_{\mathrm{gross}}(1)-\rho\Delta'(1) < 0.
\]

\begin{proof}[Proof of Proposition]
The objective is to maximize $\Phi(\gamma, \rho) = \mathrm{SR}_{\mathrm{gross}}(\gamma) - \rho \Delta(\gamma)$. 
Assume an interior maximizer $\gamma^\star$ exists and satisfies the second-order condition
$\frac{\partial^2 \Phi}{\partial \gamma^2}(\gamma^\star,\rho) < 0$.
The first-order condition (FOC) for an interior maximum is:
\begin{equation}
    \frac{\partial \Phi}{\partial \gamma} = \mathrm{SR}'_{\mathrm{gross}}(\gamma) - \rho \Delta'(\gamma) = 0.
\end{equation}
To determine how the optimal $\gamma^\star$ responds to changes in the ensemble factor $\rho$, we apply the \textbf{Implicit Function Theorem}. 
Let
\[
F(\gamma, \rho) = \mathrm{SR}'_{\mathrm{gross}}(\gamma) - \rho \Delta'(\gamma).
\]
Then, locally around $(\gamma^\star,\rho)$,
\begin{equation}
    \frac{d \gamma^\star}{d \rho} = - \frac{\partial F / \partial \rho}{\partial F / \partial \gamma}.
\end{equation}
Analyzing the components:
\begin{itemize}
    \item Numerator: $\frac{\partial F}{\partial \rho} = -\Delta'(\gamma)$. Since turnover decreases with penalty ($\Delta'(\gamma) < 0$), this term is strictly positive.
    \item Denominator: $\frac{\partial F}{\partial \gamma} = \mathrm{SR}''_{\mathrm{gross}}(\gamma) - \rho \Delta''(\gamma)$. 
    Given $\mathrm{SR}''_{\mathrm{gross}}(\gamma) < 0$ and $\Delta''(\gamma) > 0$ with $\rho \ge 0$, the entire term is strictly negative (satisfying the second-order condition for a maximum).
\end{itemize}
Substituting these signs into the sensitivity equation yields $\frac{d \gamma^\star}{d \rho} > 0$. 

Since $\rho(K)$ is a decreasing function of $K$ (ensembling reduces realized turnover costs), by the chain rule we obtain:
\begin{equation}
    \frac{d \gamma^\star}{d K} = \frac{d \gamma^\star}{d \rho} \cdot \rho'(K) < 0.
\end{equation}
This proves that as the ensemble size increases, the optimal explicit regularization parameter $\gamma^\star$ decreases.
\end{proof}

\section{Exact Gradient Accumulation for Non-Separable Objectives}\label{app:grad_accum}

Training Sharpe-ratio-maximizing portfolio models often requires a large \emph{effective} batch size, but GPU
memory limits force microbatching. Unlike additive objectives (e.g., $\sum_i \ell_i$), Sharpe-style losses are
\emph{non-separable} because their gradients depend on \emph{global} batch statistics. Consequently, naive
microbatch gradient accumulation (computing Sharpe per microbatch and summing gradients) generally does \emph{not}
equal the full-batch gradient.

We therefore use an \emph{Exact Two-Pass Microbatching} protocol: a first pass computes all global/sufficient
statistics \emph{without} storing activations, and a second pass replays the forward pass and injects the
\emph{analyatical} upstream gradient $\nabla_{r}\mathcal{L}$ into autograd. This yields the same parameter gradient
as a single full-batch backward pass under the conditions stated below, while keeping activation memory constant.

\subsection{Notation and Objectives}

Let the model output net portfolio returns arranged as a matrix
$R \in \mathbb{R}^{B\times L}$ with entries $R_{b,t}$ (sample $b$, time index $t$). Let
$N := BL$ be the number of return entries in the batch. Let $A$ denote the annualization factor (e.g., $A=252$).
We use $\varepsilon_\sigma>0$ for numerical stability in denominators and a variance floor
$\varepsilon_{\mathrm{var}}>0$ to prevent $\sigma\!\to\!0$ gradient blow-ups.

Define pooled (batch-level) statistics
\begin{equation}
\begin{aligned}
\mu
&= \frac{1}{N}\sum_{b=1}^B \sum_{t=1}^L R_{b,t},
\qquad
q
= \frac{1}{N}\sum_{b=1}^B \sum_{t=1}^L R_{b,t}^2, \\
\sigma
&= \sqrt{\max\!\left(q-\mu^2,\ \varepsilon_{\mathrm{var}}\right)}.
\end{aligned}
\label{eq:pooled_stats}
\end{equation}
Now, define per-sample statistics. For each sample $b$,
\begin{equation}
\begin{aligned}
\mu_b
&= \frac{1}{L}\sum_{t=1}^L R_{b,t},
\qquad
q_b
= \frac{1}{L}\sum_{t=1}^L R_{b,t}^2, \\
\sigma_b
&= \sqrt{\max\!\left(q_b-\mu_b^2,\ \varepsilon_{\mathrm{var}}\right)}.
\end{aligned}
\label{eq:per_sample_stats}
\end{equation}
In implementation we use a minimum clamp on the variance; when the clamp is active, the gradient through the
variance term is zero, ensuring the $1/\sigma$ factors remain bounded.

\paragraph{Loss.}
We optimize
\begin{equation}
\mathcal{L}
= \mathcal{L}_{\mathrm{pool}} + \lambda\,\mathcal{L}_{\mathrm{soft}},
\label{eq:total_loss}
\end{equation}
where the pooled Sharpe loss is
\begin{equation}
\mathcal{L}_{\mathrm{pool}}
= -\sqrt{A}\,
\frac{\mu}{\sigma+\varepsilon_\sigma}.
\label{eq:L_pool}
\end{equation}
The SoftMin term $\mathcal{L}_{\mathrm{soft}}$ is defined in the main text and is chosen to
\emph{maximize} the soft-min (smooth worst-case) per-sample Sharpe across groups. In the appendix we only require its
upstream gradient (below).

Define per-sample Sharpe
\begin{equation}
\mathrm{SR}_b
= \sqrt{A}\,
\frac{\mu_b}{\sigma_b+\varepsilon_\sigma}.
\label{eq:SR_b}
\end{equation}
Let the $B$ sequences be partitioned into $G$ groups of size $K$ (where $B=GK$), indexed by group $g\in\{1,\dots,G\}$ and within-group index $k\in\{1,\dots,K\}$. We map each sequence $b$ to a unique pair $(g,k)$. Consistent with the adversarial weights derived in App.~\ref{app:robust_objective}, we define the group-wise SoftMin weights as:
\begin{equation}
q^\star_{g,k} = \frac{\exp\!\left(-\mathrm{SR}_{g,k}/\tau\right)}{\sum_{j=1}^K \exp\!\left(-\mathrm{SR}_{g,j}/\tau\right)}.
\label{eq:pi_def}
\end{equation}
For the (negative) soft-min objective used to maximize worst-case Sharpe, the upstream gradient is:
\begin{equation}
\frac{\partial \mathcal{L}_{\mathrm{soft}}}{\partial \mathrm{SR}_{g,k}} = -\frac{1}{G}\,q^\star_{g,k}.
\label{eq:dLsoft_dSR}
\end{equation}

\subsection{Analyatical Gradients}\label{app:analyatical_sharpe_grads}

Differentiating \eqref{eq:L_pool} w.r.t.\ $R_{b,t}$ gives
\begin{equation}
\begin{aligned}
\frac{\partial \mathcal{L}_{\mathrm{pool}}}{\partial R_{b,t}}
&= -\sqrt{A}\Biggl[
\frac{1}{N(\sigma+\varepsilon_\sigma)}
\\
&\qquad\qquad
- \frac{\mu\,(R_{b,t}-\mu)}{N\,(\sigma+\varepsilon_\sigma)^2\,\sigma}
\Biggr].
\end{aligned}
\label{eq:d_pooled}
\end{equation}

Differentiating \eqref{eq:SR_b} yields
\begin{equation}
\begin{aligned}
\frac{\partial \mathrm{SR}_b}{\partial R_{b,t}}
&= \sqrt{A}\Biggl[
\frac{1}{L(\sigma_b+\varepsilon_\sigma)}
\\
&\qquad\qquad
- \frac{\mu_b\,(R_{b,t}-\mu_b)}{L\,(\sigma_b+\varepsilon_\sigma)^2\,\sigma_b}
\Biggr].
\end{aligned}
\label{eq:d_sr_per_sample}
\end{equation}

Using \eqref{eq:dLsoft_dSR} and the fact that $\mathrm{SR}_{g,k}$ depends only on its own sample,
\begin{equation}
\begin{aligned}
\frac{\partial \mathcal{L}_{\mathrm{soft}}}{\partial R_{b,t}}
&= -\frac{1}{G}\,\pi_{g(b),k(b)}\;
\frac{\partial \mathrm{SR}_b}{\partial R_{b,t}}.
\end{aligned}
\label{eq:d_soft_compact}
\end{equation}

The total upstream gradient, used for injection in Pass 2, is
\begin{equation}
\begin{aligned}
G_{b,t}
= \frac{\partial \mathcal{L}}{\partial R_{b,t}}
= \frac{\partial \mathcal{L}_{\mathrm{pool}}}{\partial R_{b,t}}
+ \lambda\,
\frac{\partial \mathcal{L}_{\mathrm{soft}}}{\partial R_{b,t}}.
\end{aligned}
\label{eq:G_def}
\end{equation}
For implementation via autograd, we aggregate the element-wise analyatical gradients into a microbatch-level tensor. Let $\mathcal{B}_m$ denote the set of indices for the samples contained in microbatch $m$. We define the injected gradient tensor $G_m \in \mathbb{R}^{| \mathcal{B}_m | \times L}$ such that its entries correspond to the scalar gradients $G_{b,t}$ derived in \eqref{eq:G_def} for all $b \in \mathcal{B}_m$ and $t \in \{1, \dots, L\}$. This tensor is then passed directly to the model's output during the second pass.

\subsection{Exact Two-Pass Microbatching Algorithm}

The procedure splits the training step into two distinct passes over the data to circumvent memory bottlenecks while maintaining mathematical exactness.

\textbf{Phase 1: Statistical Accumulation.} The model processes the data in microbatches with autograd disabled. Since gradients are not tracked, memory usage is minimal. We aggregate the sufficient statistics—sum of returns ($\sum R$) and sum of squared returns ($\sum R^2$)—globally and per sample. These aggregates allow us to compute the exact batch-wide mean $\mu$ and volatility $\sigma$, which are required to define the true gradients.

\textbf{Phase 2: Gradient Injection.} The model processes the data a second time with autograd enabled. Crucially, we do not compute the loss function forward. Instead, we use the pre-computed $\mu$ and $\sigma$ from Phase 1 to evaluate the analyatical gradient formulas derived in Section~\ref{app:analyatical_sharpe_grads}. This yields a gradient tensor $G_m$ for the microbatch outputs. We inject this tensor directly into the backward pass using \texttt{r\_m.backward(gradient=G\_m)}. This propagates the exact gradient of the full-batch Sharpe ratio through the computation graph of the microbatch.

\begin{algorithm}[t]
\caption{Exact Two-Pass Microbatching for Non-Separable Sharpe Objectives}
\label{alg:exact_two_pass}
\begin{algorithmic}[1]
\State \textbf{Inputs:} dataset $\mathcal{D}$, microbatch size $M$, model parameters $\theta$
\State \textbf{Initialize} pooled sums $S_1 \leftarrow 0$, $S_2 \leftarrow 0$
\State \textbf{Initialize} per-sample sums $\{S_{1,b}\}_{b=1}^B \leftarrow 0$, $\{S_{2,b}\}_{b=1}^B \leftarrow 0$
\State \textbf{// Pass 1: Collect Statistics (No Activations)}
\For{microbatch $m$ in $\mathcal{D}$}
  \State \textbf{Set deterministic RNG seed} for microbatch $m$
  \State $R_m \leftarrow \text{Model}_\theta(\text{input}_m)$ \Comment{forward only}
  \State Accumulate pooled: $S_1 \mathrel{+}= \sum R_m$, \ $S_2 \mathrel{+}= \sum R_m^2$
  \State Accumulate per-sample: $S_{1,b} \mathrel{+}= \sum_t R_{b,t}$, \ $S_{2,b} \mathrel{+}= \sum_t R_{b,t}^2$
\EndFor
\State Compute $\mu, q, \sigma$ from $S_1,S_2$ via \eqref{eq:pooled_stats}
\State Compute $\mu_b, q_b, \sigma_b$ from $S_{1,b},S_{2,b}$ via \eqref{eq:per_sample_stats}
\State Compute all $\mathrm{SR}_b$ via \eqref{eq:SR_b} and SoftMin weights $\pi_{g,k}$ via \eqref{eq:pi_def}

\State \textbf{// Pass 2: Inject analyatical Upstream Gradient}
\For{microbatch $m$ in $\mathcal{D}$}
  \State \textbf{Reset deterministic RNG seed} for microbatch $m$
  \State $R_m \leftarrow \text{Model}_\theta(\text{input}_m)$ \Comment{forward with autograd}
  \State Build $G_m$ using \eqref{eq:d_pooled}, \eqref{eq:d_sr_per_sample}, \eqref{eq:d_soft_compact}, \eqref{eq:G_def}
  \State $R_m.\text{backward}(G_m)$ \Comment{inject upstream gradient}
\EndFor
\State $\text{Optimizer.step}()$
\end{algorithmic}
\end{algorithm}

We must consider some exactness conditions. The two-pass protocol reproduces the same parameter gradient as a single full-batch backward pass \emph{provided that}
the per-sample forward mapping is identical across passes and independent of microbatch composition. Concretely:
(i) \textbf{No batch-coupled layers:} the network must not contain layers whose outputs depend on microbatch
statistics (e.g., BatchNorm) or on other samples in the microbatch.
(ii) \textbf{No stateful forward updates:} modules that update running buffers during the forward pass must be
disabled/frozen.
(iii) \textbf{Identical data order and grouping:} Pass 1 and Pass 2 must iterate over the same samples in the same
order, and the SoftMin grouping/indexing used to compute $\pi_{g,k}$ must match the mapping used when injecting
gradients in Pass 2 (i.e., no reshuffling between passes).
(iv) \textbf{Deterministic stochastic layers:} any stochasticity (e.g., Dropout) must be controlled so that $R_m$ is
identical in both passes (handled by per-microbatch RNG reseeding above).

\section{Analysis of the Robust Objective}
\label{app:robust_objective}

This appendix provides a formal interpretation of the windowed \emph{SoftMin} aggregation used in
Sec.~\ref{sec:objective:robust_loss}. We show that the resulting objective admits (i) a KL-penalized
distributionally robust optimization (DRO) interpretation via the Gibbs/Donsker--Varadhan variational
principle, and (ii) a connection to the coherent risk measure \emph{Entropic Value-at-Risk (EVaR)}.

\subsection{SoftMin as KL-penalized DRO (Variational Form)}
\label{app:robust:dv}

We consider $B$ training windows and their realized Sharpe ratios $\{\mathrm{SR}_b\}_{b=1}^B$.
Define window \emph{losses} $Z_b(\theta) \coloneqq -\mathrm{SR}_b(\theta)$, and let $P$ denote the
uniform empirical distribution over windows ($P(b)=1/B$). For a temperature $\tau>0$, our SoftMin-in-Sharpe
objective is equivalently a softmax in the losses:
\begin{equation}
\begin{aligned}
\mathcal{L}_{\text{soft}}(\theta)
&\!=\! \tau \log\!\left(\frac{1}{B}\sum_{b=1}^B \exp\!\left(\frac{Z_b(\theta)}{\tau}\right)\right)
\!=\! \tau \log \mathbb{E}_{P}\!\!\left[\exp\!\left(\frac{Z}{\tau}\right)\right]\!.
\end{aligned}
\label{eq:softmax_entropic_app}
\end{equation}

As $\tau\to0^+$, $\mathcal{L}_{\text{soft}}(\theta)\to \max_b Z_b(\theta)= -\min_b \mathrm{SR}_b(\theta)$,
recovering a worst-window (min-Sharpe) criterion; as $\tau\to\infty$, it approaches the average loss.

\begin{proposition}[Gibbs / Donsker--Varadhan variational principle (discrete)]
\label{prop:dv_discrete}
The entropic aggregate in \eqref{eq:softmax_entropic_app} admits the dual representation
\begin{equation}
\tau \log \mathbb{E}_{P}\!\left[\mathrm{e}^{Z/\tau}\right]
= \sup_{Q \ll P}
\left\{
\mathbb{E}_{Q}[Z] - \tau \,\mathrm{KL}(Q\|P)
\right\},
\end{equation}
where $Q$ is an adversarial reweighting over windows and $Q\ll P$ is automatic here since $P$ is uniform.
\end{proposition}

\begin{proof}
The duality follows directly from the variational representation of the log-sum-exp function, a standard result in convex analysis \citep{boucheron2013concentration}. For completeness, we derive it for the discrete case below. Let $q_b$ be the weights of $Q$ with $\sum_{b=1}^B q_b=1$. Since $P(b)=1/B$,
\(
\mathrm{KL}(Q\|P)=\sum_{b=1}^B q_b \log\!\big(q_b/(1/B)\big)=\sum_{b=1}^B q_b \log(B q_b).
\)
Consider the Lagrangian
\begin{equation}
J(q,\eta)
= \sum_{b=1}^B q_b Z_b
- \tau \sum_{b=1}^B q_b \log(B q_b)
+ \eta\!\left(1-\sum_{b=1}^B q_b\right).
\end{equation}
Stationarity gives
\(
0=\partial J/\partial q_b = Z_b - \tau\big(1+\log(B q_b)\big)-\eta,
\)
hence
\begin{equation}
q_b^\star
= \frac{\exp(Z_b/\tau)}{\sum_{j=1}^B \exp(Z_j/\tau)}.
\label{eq:adversarial_weights}
\end{equation}
Substituting $q^\star$ into $\sum_b q_b Z_b - \tau\sum_b q_b \log(B q_b)$ yields
$\tau \log\!\big(\frac{1}{B}\sum_{b=1}^B \mathrm{e}^{Z_b/\tau}\big)$, i.e.\ \eqref{eq:softmax_entropic_app}.
\end{proof}

The weights $q_b^\star$ in \eqref{eq:adversarial_weights} provide a diagnostic: concentrated mass on a
small set of windows indicates that those regimes dominate the gradient signal and are the current
``stress scenarios'' for the model.

\subsection{Connection to KL-ball DRO and EVaR}
\label{app:robust:evar}

A ``hard'' DRO variant constrains the adversary to a KL-divergence ball of radius $C$:
\begin{equation}
\mathcal{R}_C(Z)
= \sup_{Q:\,\mathrm{KL}(Q\|P)\le C} \mathbb{E}_Q[Z].
\end{equation}
By Lagrangian duality together with Proposition~\ref{prop:dv_discrete}, this is equivalent to
\begin{equation}
\mathcal{R}_C(Z)
= \inf_{\lambda>0}
\left\{
\lambda \log \mathbb{E}_P\!\left[\mathrm{e}^{Z/\lambda}\right]
+ \lambda C
\right\}.
\label{eq:kl_ball_dual}
\end{equation}

Choosing the divergence budget $C=-\log(1-\alpha)$ yields \emph{Entropic Value-at-Risk (EVaR)} at confidence
level $\alpha\in(0,1)$:
\begin{equation}
\mathrm{EVaR}_\alpha(Z)
= \inf_{\lambda>0}
\left\{
\lambda \log \mathbb{E}\!\left[\mathrm{e}^{Z/\lambda}\right]
- \lambda \log(1-\alpha)
\right\}.
\label{eq:evar_def}
\end{equation}

In DeePM, the training temperature $\tau$ can be viewed as selecting a particular member of this
entropic/EVaR family (i.e., evaluating \eqref{eq:evar_def} at $\lambda=\tau$, up to the additive constant
$-\tau\log(1-\alpha)$ when $\alpha$ is fixed). Thus SoftMin training implements a smooth, tail-sensitive
surrogate that emphasizes poor historical windows, while remaining differentiable and stable.


\section{Experimental Implementation Details}\label{app:exp_details}

This appendix provides the granular specifications required to reproduce the experimental results, including hyperparameter search spaces, and the specific optimization protocols employed.

\subsection{Optimization and Training Dynamics}

We optimize the parameters using \textbf{AdamW} \citep{loshchilov2019adamw} to decouple weight decay from gradient updates, which is crucial for stabilizing Transformers. We use a fixed learning rate of $\eta = 10^{-4}$ without annealing, as the non-stationary nature of financial data often makes cyclic or decaying schedules suboptimal for continuous learning.

Given the low signal-to-noise ratio in financial time series, validation metrics can be highly volatile across epochs. To prevent premature stopping due to lucky/unlucky noise, we implement an \textbf{Exponential Moving Average (EMA)} on the validation metric (Sharpe Ratio). 
Let $S_k^{\text{val}}$ be the raw validation Sharpe at iteration $k$. The smoothed metric $\tilde{S}_k$ updates as:
\begin{equation}
    \tilde{S}_k = \alpha_{\text{smooth}} S_k^{\text{val}} + (1 - \alpha_{\text{smooth}}) \tilde{S}_{k-1},
\end{equation}
with $\alpha_{\text{smooth}} = 0.45$. Training terminates if $\tilde{S}_k$ fails to improve by $\delta_{\text{min}}=0.001$ for a patience defined per architecture (see Table \ref{tab:hyperparams}). We employ a ``burn-in" of 20 iterations before enabling the early stopping trigger to allow the optimizer to stabilize.

Models were trained on a single NVIDIA GeForce RTX 3090 (24GB) GPU. 

\subsection{Existence Masking and Missingness Handling}\label{sec:method:masking}
Since the architecture is permutation equivariant, it inherently handles variable-sized sets of assets ($N_t \le N$). However, for practical training on GPUs requiring fixed tensor shapes, we construct the data tensor on a complete $(i,t)$ grid with forward-filling.
To prevent learning from synthetically filled values when a contract is not present, DeePM uses: (i) an existence indicator appended to dynamic inputs, (ii) a key-padding mask $\mathrm{kpm}_\ell$ in cross-sectional attention, and (iii) explicit zeroing of invalid node rows after spatial blocks. Finally, positions are masked via $m_{i,t}$ in Eq.~\eqref{eq:net_return}.

\subsection{Hyperparameter Search Space}
We perform a random grid search over the key architectural and regularization parameters. For the final ensemble, we select the top seeds from 100 random trials based on the best smoothed validation Sharpe ratio. Table \ref{tab:hyperparams} details the search grid for the proposed DeePM model versus the Momentum Transformer baseline.

\begin{table}[h]
    \centering
    \small
    \caption{Hyperparameter Search Space (DeePM vs. Baseline)}
    \label{tab:hyperparams}
    \begin{tabular}{lll}
        \toprule
        \textbf{Parameter} & \textbf{DeePM (Proposed)} & \textbf{Momentum Transformer} \\
        \midrule
        \multicolumn{3}{l}{\underline{\textit{Architecture}}} \\
        Hidden Dimension ($d_{\text{model}}$) & $\{64, 128\}$ & $\{64, 128\}$ \\
        Attention Heads & $\{2, 4\}$ & $4$ \\
        Dropout & $\{0.3, 0.4, 0.5\}$ & $\{0.3, 0.4, 0.\}$ \\
        \midrule
        \multicolumn{3}{l}{\underline{\textit{Optimization}}} \\
        Batch Size & $64$ & $\{64, 128\}$ \\
        Learning Rate & $10^{-4}$ & $\{10^{-4}, 3 \cdot 10^{-4}\}$ \\
        Weight Decay & $\{10^{-5}, 10^{-4}, 10^{-3}\}$ & $\{10^{-5}, 10^{-4}, 10^{-3}\}$ \\
        Max Gradient Norm & $\{0.5, 1.0\}$ & $\{0.5, 1.0\}$ \\
        Training Burn-in Steps & 50 & 5 \\
        SoftMin Temp ($\tau$) & 0.2 & - \\
        SoftMin Scalar ($\lambda$) & 0.1 & - \\
        \midrule
        \multicolumn{3}{l}{\underline{\textit{Training Config}}} \\
        Sequence Length & $84$ days & $84$ days \\
        Burn-in Steps & $21$ days & $21$ days \\
        Iterations & $1000$ & $200$ \\
        Early Stopping Patience & $50$ & $25$ \\
        Top $N$ Ensemble & $25$ & $25$ \\
        \bottomrule
    \end{tabular}
\end{table}

\end{document}